\documentclass[aps,notitlepage,nofootinbib,longbibliography,twocolumn,superscriptaddress,10pt]{revtex4-1}
\usepackage{amsmath}
\usepackage{amsthm, amssymb}
\usepackage{slashed} 
\usepackage{graphicx}
\usepackage{xcolor}
\usepackage{tikz}
\usetikzlibrary{calc,fadings,decorations.pathreplacing,shapes,shapes.multipart,arrows,shapes.misc,intersections,positioning,patterns}
\usepackage{bm}
\usepackage[colorlinks=true,citecolor=blue,linkcolor=magenta]{hyperref}
\usepackage{dsfont}
\usepackage{changepage}
\usepackage{array}

\newcommand{\bit}{\begin{itemize}}
\newcommand{\eit}{\end{itemize}}

\newcommand{\f}{\frac}
\renewcommand{\>}{\right\rangle}
\newcommand{\<}{\left\langle}
\newcommand{\ba}{\begin{align}}
\newcommand{\ea}{\end{align}}
\newcommand{\be}{\begin{equation}}
\newcommand{\ee}{\end{equation}}
\newcommand{\bi}{\begin{itemize}}
\newcommand{\ei}{\end{itemize}}
\newcommand{\lf}{\left(}
\newcommand{\ri}{\right)}
\newcommand{\dd}{\mathrm{d}}

\newcommand{\nccp}{\mathrm{NCCP}}

\renewcommand{\vec}[1]{\boldsymbol{\mathbf{#1}}}

\graphicspath{{}{./images/}}

\begin{document}
\title{Emergence and spontaneous breaking of approximate $\mathrm{O}(4)$ symmetry \\ at a weakly first-order deconfined phase transition}

\author{Pablo Serna}
\affiliation{Laboratoire de Physique de l'Ecole Normale Sup\'erieure, ENS, Universit\'e PSL, CNRS, Universit\'e Paris-Diderot, Sorbonne Universit\'e, Sorbonne Paris Cit\'e, 24 rue Lhomond, 75005 Paris, France}
\author{Adam Nahum}
\affiliation{Theoretical Physics, Oxford University, 1 Keble Road, Oxford OX1 3NP, United Kingdom}

\date{\today}

\begin{abstract}
We investigate approximate emergent nonabelian symmetry in a class of weakly first order `deconfined' phase transitions using Monte Carlo simulations and a renormalization group analysis.
We study a  transition in a 3D classical loop model that is analogous to a deconfined 2+1D quantum phase transition in a magnet with reduced  lattice symmetry. The transition is between  the N\'eel phase and a \textit{twofold} degenerate valence bond solid (lattice-symmetry-breaking) phase.
The combined order parameter at the transition is effectively a four-component superspin.
It has been argued that in some weakly first order `pseudocritical' deconfined phase transitions, the renormalization group flow can take the system very close to the \textit{ordered} fixed point of the symmetric $O(N)$ sigma model,
where $N$ is the total number of `soft' order parameter components,
 despite the fact that $O(N)$ is not a microscopic symmetry. This yields a first order transition with unconventional phenomenology.
We argue that this occurs in the present model, with $N=4$.
This means that there is a regime of lengthscales in which the transition resembles a `spin-flop' transition in the {ordered} $O(4)$ sigma model.
We give numerical evidence for (i) the first order nature of the transition, 
(ii) the emergence of $O(4)$ symmetry to an accurate approximation, and 
(iii) the existence of a regime in which the emergent $O(4)$ is `spontaneously broken', with distinctive features in the order parameter probability distribution.
These results may be relevant for other models studied in the  literature, including 2+1D QED with two flavours, the `easy-plane' deconfined critical point, and the N\'eel--VBS transition on the rectangular lattice.
\end{abstract}

\maketitle

\tableofcontents

\section{Introduction}
\label{sec:intro}

The phenomenology of deconfined criticality \cite{deccp,deccplong} in models with small `$N$' --- for example  spin-1/2 magnets with various symmetries in 2+1D 
\cite{SandvikJQ,melkokaulfan,lousandvikkawashima,Banerjeeetal,
Sandviklogs,Kawashimadeconfinedcriticality,Jiangetal,deconfinedcriticalityflowJQ,kukloveasyplane,kuklovetalDCPSU(2),
kaulsandviklargen,sandvik2parameter,kauleasyplane1,qin2017duality,zhang2018continuous}, and related classical models \cite{DCPscalingviolations,emergentso5,sreejith2018emergent,powellmonopole} --- has turned out  to be more interesting and more subtle than could have been guessed from simple large $N$ approximations.
Many phenomena have been invoked in order to explain perplexing numerical results: we are gradually elucidating which  are only mirages, and which are the crucial features of the problem.
Here we demonstrate a remarkable weakly first-order `deconfined' phase transition, with unusual phenomenology, in a model with two competing order parameters. This phase transition combines the ingredient of emergent symmetry with that of `quasiuniversality' \cite{DCPscalingviolations,wang2017deconfined}  at weakly first order deconfined transitions.

Emergent nonabelian symmetries,  which unite distinct order parameters,
are a key concept for understanding deconfined criticality in small `$N$' models \cite{tsmpaf06,tanakahu,emergentso5,wang2017deconfined}.
At the N\'eel to valence-bond-solid (N\'eel--VBS) transition in square lattice spin-1/2 magnets, with full spin rotation symmetry, five order parameter components become `soft' at the transition. These are the three components of the N\'eel vector $\vec N$ and the two components of the VBS order parameter $\vec \phi$.
Numerics show a very accurate $SO(5)$ symmetry uniting these five components at the transition \cite{emergentso5,sreejith2018emergent,SandvikSpectrum}. We may write them as a superspin,
\be\label{5cptvector}
{\bf n} = (N_x, N_y, N_z, \phi_x, \phi_y).
\ee
At currently accessible system sizes, the $SO(5)$ symmetry for $\vec n$ looks like exact symmetry of the infra-red theory, which becomes increasingly accurate at longer lengthscales. At asymptotically long lengthscales it may well only  be an approximate symmetry (see \cite{wang2017deconfined,sreejith2018emergent,PolandReview} for discussions), but it seems to be a robust feature of models described by the  noncompact $\mathrm{CP}^1$ field theory ($\nccp^1$ \cite{motrunichvishwanath1})  originally proposed for the Neel-VBS transition \cite{deccp}. For example,
$SO(5)$ also emerges \cite{sreejith2018emergent} in a classical dimer model that has very different microscopic symmetries to the N\'eel--VBS models \cite{PowellChalkerPRL,CharrierAletPujol,Chenetal,PowellChalkerPRB}. 
Emergent $SO(5)$ is equivalent to an infra-red self-duality of the $\nccp^1$ field theory, and it also opens up the possibility of relationships with other field theories, including two-flavour quantum chromodynamics \cite{wang2017deconfined}. A variant of the latter has recently been proposed to describe a confinement transition with emergent $SO(5)$ \cite{gazit2018confinement}.

A different class of putative `deconfined' phase transitions are those involving only \textit{four} soft order parameter components. 
The best-studied example is the `easy-plane' N\'eel--VBS transition, with two N\'eel and two VBS components \cite{kukloveasyplane, kragseteasyplane, kauleasyplane1, kauleasyplane2,qin2017duality,zhang2018continuous, motrunichvishwanath1, motrunichvishwanath2}
\be\label{4cptvector1}
{\bf n} = (N_x, N_y, \phi_x, \phi_y).
\ee
It was suggested that this transition, if continuous, may have emergent $O(4)$ symmetry \cite{wang2017deconfined,qin2017duality}.
There are models where the transition is clearly first order \cite{kukloveasyplane, kragseteasyplane, kauleasyplane1, kauleasyplane2,
Chenetal} (including a self-dual  \cite{motrunichvishwanath1} lattice model \cite{geraedtsmotrunich1,geraedtsmotrunich2}), but  it has recently been claimed that the transition is continuous in other models \cite{qin2017duality,zhang2018continuous}.
It was also pointed out recently \cite{metlitski2017intrinsic, sato2017dirac} that the same $O(4)$--invariant long-distance theory may well apply to another set of transitions: these have full ${SO}(3)$ spin rotation symmetry, but only a \textit{one}-component VBS order parameter \cite{tsmpaf06}:
\be\label{eq:superspin}
{\bf n} = (N_x, N_y, N_z, \varphi).
\ee
In this last set of examples there is, loosely speaking, a semi-microscopic $SO(3)\times \mathbb{Z}_2$ symmetry, where the $\mathbb{Z}_2$ really refers to lattice symmetries (e.g. translation) that change the sign of the VBS order parameter $\varphi$.

One member of this class is the N\'eel--VBS transition on a square lattice with rectangular anisotropy, which suppresses one of the two possible directions for the columnar VBS order parameter \cite{metlitski2017intrinsic,tsmpaf06}. 
Another is a N\'eel to (twofold degenerate) VBS transition that has been studied in a  fermionic model on the honeycomb lattice: numerics support emergent $O(4)$ symmetry here \cite{sato2017dirac}.
A third example is a transition  \cite{metlitski2017intrinsic} for spin-1s on the isotropic square lattice, between the N\'eel phase and a VBS phase \cite{WangNematicity2015,KomargodskiWalls2018} with two degenerate ground states.

The 3D classical transition we study below is analogous to a 2+1D  N\'eel--VBS transition for spin-1/2s on a square lattice with reduced lattice symmetry, and also has a four-component order parameter of the form in Eq.~\ref{eq:superspin}, with $SO(3)\times{\mathbb{Z}_2}$ symmetry.

We can think of the  four-component cases (\ref{4cptvector1}) and (\ref{eq:superspin}) as descending from the five-component case under deformations that disfavour one of the components of (\ref{5cptvector}).
It is worth noting that the relation between these deformations is highly non-obvious in the usual $\nccp^1$ language. 
Easy-plane anisotropy and, say, rectangular anisotropy \cite{metlitski2017intrinsic} correspond to deformations of very different types that reduce the symmetry of the $\nccp^1$ model in different ways (easy plane anisotropy yields a quartic potential for the spinon field, while rectangular anisotropy gives a fugacity for strength-two Dirac monopoles).\footnote{Close to the critical/pseudocritical point of the original NCCP$^1$ model both of these perturbations are relevant, and they are related by emergent $SO(5)$. Note  however  that here we are considering deformations of $O(1)$ strength, rather than infinitesimal perturbations.}
However if $O(4)$ symmetry emerges there is the possibility that these various models converge under RG flow. Even more strikingly, the easy-plane model has been conjectured to be dual to  two-flavour quantum electrodynamics ($N_f=2$ QED) \cite{karchtong,wang2017deconfined}, and this theory may also acquire $O(4)$ in the infrared. 
A conjectured self-duality of $N_f=2$ QED  \cite{qeddual,karchtong,seiberg2, JianEmergent2017} would imply $O(4)$ in that theory  (which may also be argued for heuristically by a mapping to the $O(4)$ sigma model at $\theta=\pi$ \cite{tsmpaf06}, and which has some support from simulations \cite{KarthikFlavor2017}).

A second useful concept is that of pseudocriticality. 
This is a regime of `slow' renormalization group (RG) flow, associated with a more or less well-defined \textit{line} in coupling constant space that attracts nearby flow lines, leading to `quasiuniversal' behaviour that is only weakly dependent on the bare parameters \cite{wang2017deconfined}.
This phenomenon can be made precise in various field theories with a parameter that does not flow under the RG, and  was proposed as an explanation for various puzzling phenomena at deconfined phase transitions in \cite{DCPscalingviolations,wang2017deconfined}.
A classic example of this phenomenon is the Potts model, where as the number of states $Q$ is increased the continuous transition gives way to a first-order one at  a universal critical value \cite{baxterpotts,nienhuispotts,nauenbergscalapino,cardynauenbergscalapino,IinoDetecting2018}. This is due to the annihilation of a critical and a tricritical fixed point. (Fixed point annihilation has also been discussed in QCD and QED, among other models \cite{Gies, kaplan,braun2014phase,GiombiConformal2016,HerbutChiral2016,GukovRG2017,PhaseTransitionsCPNSigmaModel}.) For $Q$ slightly larger than the critical value, there is a first order transition with a parametrically large correlation length and quasiuniversal properties that depend only parametrically weakly \cite{wang2017deconfined} on the bare couplings. (See \cite{IinoDetecting2018} for a recent numerical study of  the Potts case.)

In models that show a pseudocritical regime, the simplest possibility is that at the very longest lengthscales the RG flow is to a discontinuity fixed point, i.e. a first order transition.
However things can be more interesting if an (approximate) emergent symmetry is established during the quasiuniversal RG flow associated with pseudocriticality, i.e. if the attractive flow line has the higher symmetry  \cite{wang2017deconfined}.
In this case the emergent symmetry persists even into the coexistence regime --- i.e. the regime of lengthscales where the order parameters no longer appear to be scaling to zero with system size (as at a critical point) but have instead saturated to a finite value. This gives an unconventional first-order transition which resembles a spin-flop transition in an \textit{ordered} sigma model with the higher symmetry (i.e., a transition driven by changing the sign of an anisotropy in the sigma model). We review this RG picture below.

It is plausible that the above scenario is what ultimately happens in the models that show an emergent $SO(5)$ symmetry. But if so, the lengthscale required to see it seems to be inaccessible at present: we can access the `pseudocritical' regime but not the `spin flop' regime at larger scales. (It is also conceivable that the ultimate fate of the models with $SO(5)$ symmetry is something else, though this is constrained by conformal bootstrap \cite{SimmonsDuffinSO(5),Nakayama,PolandReview}.)

By contrast, we show here that the phenomenon of a weakly first order transition with emergent approximate $O(4)$ symmetry \textit{can} be seen in models for deconfined criticality with \textit{four} order parameter components, or rather in at least one such model.
In the model we study the lengthscale associated with the first order transition is long enough to allow $O(4)$ to emerge with good accuracy (we demonstrate this directly using the order parameter distribution) and short enough that we can convincingly demonstrate that the transition is first-order. 

The resulting first-order transition has an interesting phenomenology, because the two competing phases, which coexist at the critical coupling, are related by an effective continuous symmetry in the appropriate range of lengthscales. This has characteristic signatures in the probability distribution of the order parameters, which takes a simple universal form, and of quantities like the energy, which do \textit{not} have the double peaked shape familiar in conventional first order transitions.
The model we study is a classical 3D loop model \cite{DCPscalingviolations}  that is closely related to a deconfined critical point for 2+1D square lattice magnets, but with reduced lattice symmetry (the loop model can be viewed as a deformation of the partition function of a quantum magnet to impose isotropy in the three `spacetime' dimensions, making the model more convenient for simulations).
We argue that this model shows the basic features of the RG scenario outlined above.

Work on field theoretic  dualities has recently revealed unexpected connections between 2+1D field theories, and  as a result 
the present results are relevant to a variety of other models.
Various examples of deconfined criticality with four order-parameter components have been discussed in the literature, and as mentioned above
the easy-plane transition has been argued to be dual to 2+1D quantum electrodynamics with $N_f=2$.
It would be worth looking for the phenomena we discuss in those models.
(The status of duality webs relating different theories, in cases where the emergent symmetry implied by the dualities is approximate rather than exact, is discussed in \cite{wang2017deconfined}.)
Recent numerics has however argued for a continuous  easy-plane transition rather than a weakly first-order one \cite{qin2017duality,zhang2018continuous} (see \cite{kauleasyplane2,JianEmergent2017} for discussions of relations between different easy-plane models), and for scale invariance in $N_f=2$ QED \cite{qedcft}.
In the light of the  relationships expected between the various field theories this difference with what we find here is  surprising and would be worth examining further.

\section{RG picture}
\label{sec:rgpicture}

Before describing a particular model, in this section we review the scenario of \cite{wang2017deconfined} for a weakly first order deconfined phase transition with approximate emergent symmetry. 
For concreteness, we focus on the scenario with a four-component order parameter which will be relevant to us below.
For a simplified picture, it is useful to think of the nonlinear sigma model as the effective field theory for this order parameter \cite{tsmpaf06}.
This  theory is not  precisely defined, since the sigma model is a nonrenormalizable effective field theory whose  definition is regularization dependent, but this will not matter for the following qualitative discussion of the RG flows (see \cite{wang2017deconfined} for a more careful discussion).

The sigma model for the four-component field ${\bf n}$, with ${\bf n^2}=1$, is 
\be
\label{eq:sigmamodel}
\mathcal{L} = 
\f{1}{2g} (\partial {\bf n})^2 
+ \f{i\pi \epsilon_{abcd}}{\text{Area}(S^3)}
 n_a \partial_{x_1} n_b \partial_{x_2} n_c \partial_{x_3} n_d
 + \ldots
\ee
The second term is the topological $\Theta$ term whose presence was argued for in \cite{tsmpaf06} (see also \cite{tanakahu} for a related discussion of the 5-component case).
The `$\ldots$' include  perturbations that reduce the symmetry to the microscopic symmetry of the lattice model of interest.

These perturbations can be classified into representations of $O(4)$ symmetry.
We assume  that the ones which are important for us in the pseudocritical regime (described below) --- i.e. the most relevant allowed perturbations --- are  two- and four-index symmetric traceless $O(4)$ tensors $X^{(2)}_{ab}$ and $X^{(4)}_{abcd}$ \cite{emergentso5,wang2017deconfined}.
We assume that the former is effectively \textit{relevant} in the pseudocritical regime and the latter is effectively irrelevant. At the level of symmetry ${X^{(2)}_{ab}  \sim n_a n_b - \f{1}{4} \delta_{ab}}$ and ${X^{(4)}_{abcd}  \sim n_a n_b n_c n_d - \cdots}$.

In the loop model we study below, a microscopic $SO(3)$ symmetry acts on $(n_1,n_2,n_3)=(N_x,N_y,N_z)$. There is no microscopic continuous symmetry acting on $n_4=\varphi$, but this field changes sign under lattice symmetries, for example appropriate lattice translations. In the continuum there is effectively a $\mathbb{Z}_2$ symmetry for $n_4$.
We will refer loosely to the model as having $SO(3)\times \mathbb{Z}_2$ symmetry. This is analogous to the case of the N\'eel-VBS transition on a rectangular lattice \cite{metlitski2017intrinsic} (though with some differences, see Sec.~\ref{sec:model}).

In these models the microscopic symmetries allow the following perturbations built from the above operators. The first is the strongly \textit{relevant} anisotropy between N\'eel and VBS that drives the transition:
\ba
\delta \mathcal{L} & = - g_R X^{(2)}_{44},
&
g_R & \propto (J-J_c).
\end{align}
(Here $J$ is the coupling that tunes the transition in the microscopic model.)
The second is a higher order anisotropy between N\'eel and VBS that is assumed to be effectively \textit{irrelevant} in the pseudocritical regime:
\be\label{eq:irrelevant1}
\delta \mathcal{L} = g_I X^{(4)}_{4444}.
\ee
A key point  is that $\mathbb{Z}_2\times SO(3)$ allows only one (effectively) relevant coupling at the transition \cite{metlitski2017intrinsic}.

\begin{figure}[t]
 \begin{center}
 \includegraphics[width=0.95\linewidth]{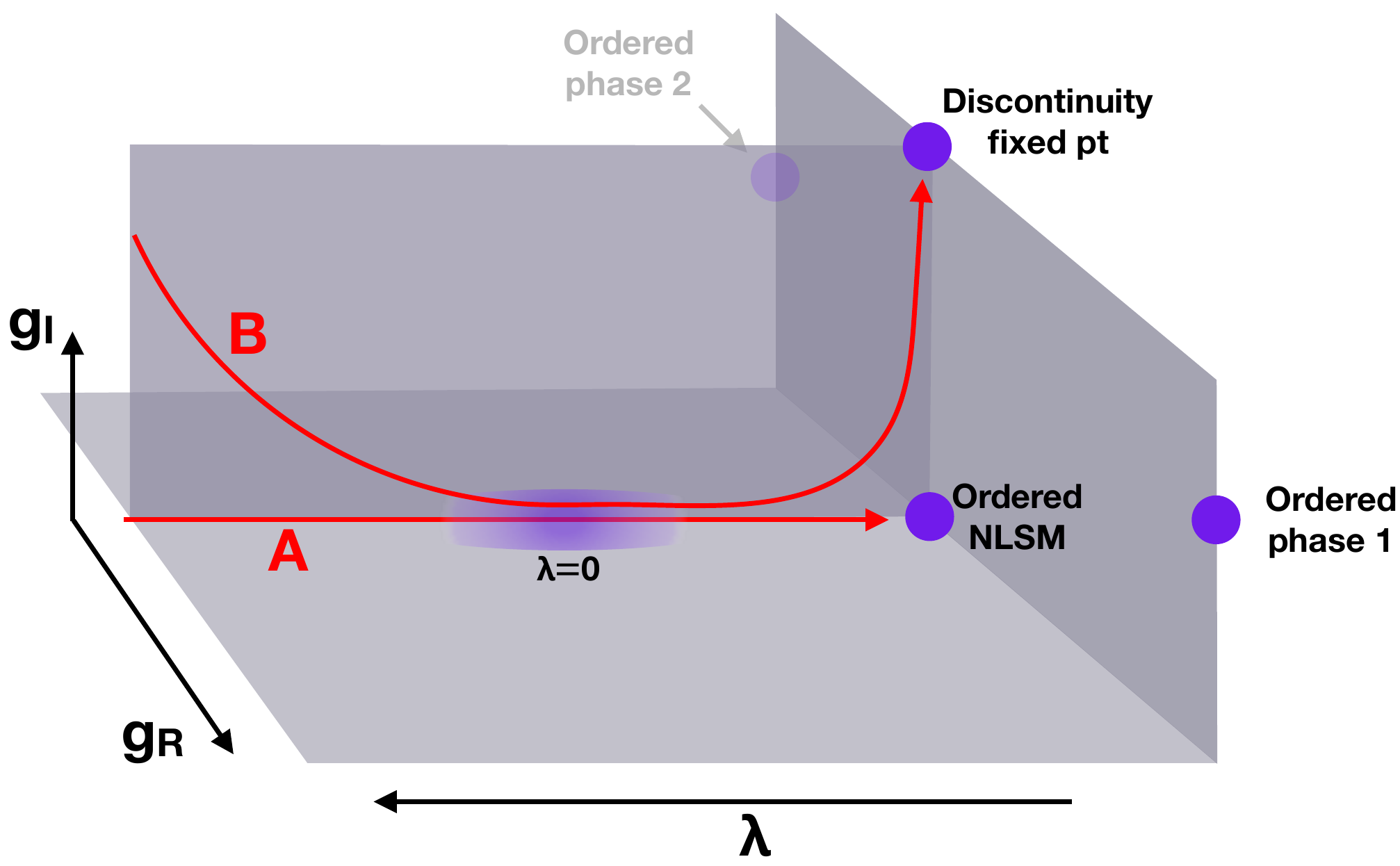}
 \end{center}
\caption{RG flows at $J_c$, i.e. with vanishing strongly relevant coupling $g_R=0$ (schematic).
Flow line {\bf A}: A hypothetical exactly $O(4)$-symmetric model is assumed to flow slowly through the pseudocritical region (blurred purple area), eventually reaching the ordered phase of the $O(4)$ sigma model. Flow line {\bf B}:
The coupling $g_I$ is effectively irrelevant in the pseudocritical regime, 
so the flow is attracted towards $g_I=0$. 
However the coupling is strongly relevant at the ordered fixed point of the nonlinear sigma model and ultimately it drives the flow to the discontinuity fixed point representing a trivial first order transition. There is an intermediate regime showing the properties of the ordered $O(4)$-symmetric sigma model.}
\label{fig:jcflows}
\end{figure}

A similar picture applies at the easy-plane N\'eel--VBS transition, despite the rather different microscopic symmetry. In easy-plane models  on the square lattice there 
is a relevant perturbation $\sum_{a=1,2} X^{(2)}_{aa}$ driving the transition and
 typically two leading irrelevant perturbations allowed by the (different) microscopic symmetry  \cite{wang2017deconfined}. These are a higher-order anisotropy between N\'eel and VBS, and fourfold symmetry breaking for the VBS:
\be\label{eq:irrelevant2}
\delta \mathcal{L} = g_I^{(1)} \sum_{a,b=1,2} X_{aabb} + g_I^{(2)} \sum_{a=3,4} X_{aaaa}.
\ee

The $O(4)$ pseudocritical regime described below may also arise in $N_f=2$ QED$_3$. An important difference in that context is that perturbations built from $X^{(2)}$ and $X^{(4)}$ are forbidden by microscopic $SU(2)$ flavour symmetry \cite{tsmpaf06,wang2017deconfined}, indicating that $O(4)$ is even more robust at long lengthscales than in the cases we discuss here.

Now let us consider how the renormalization group flows of a theory with a higher symmetry $G$ (here ${G=O(4)}$) may allow for an unconventional first order transition in a microscopic model without $G$ symmetry \cite{wang2017deconfined}.
We assume there is a flow-line in the $G$-symmetric parameter space along which the RG flow becomes very slow.
Let us parameterize the flow line by $\lambda$, with $\lambda$ decreasing under the flow, and with the slow region of RG flows being close to $\lambda=0$. 
For a heuristic picture, we can imagine that this represents the flow of the coupling constant in the nonlinear sigma model above, with $\lambda > 0$ corresponding to the region of small stiffness (large $g$) and $\lambda \rightarrow - \infty$ corresponding to the ordered fixed point at large stiffness ($g=0$).

`Slow' is a qualitative rather than a precise statement in the present context: 
its usefulness will ultimately be determined by comparing the following results with numerics.
However the idea of pseudocriticality due to slow RG flows \textit{can} be made precise in cases where we have
 a tuning parameter that controls the slowness of the RG flows. 
 One example is the $Q$-state Potts model for small $\Delta^2 \equiv Q-Q_c(d)$, where $Q_c(d)$ is the largest value of $Q$ for which the transition is continuous in $d$ dimensions \cite{baxterpotts,nienhuispotts,nauenbergscalapino,cardynauenbergscalapino,IinoDetecting2018}.
Another is the $\nccp^{n-1}$ model for small $\Delta^2\equiv d-d_c(n)$, where $d_c(n)$ is the largest value of $d$ for which the transition is continuous at a given $n$ \cite{DCPscalingviolations}. 
In these examples the RG equations can be expanded systematically in $\Delta^2$ for small $\Delta^2$. When $\Delta^2\leq 0$ the transition is continuous. $\Delta^2\gtrsim 0$ is the `pseudocritical' regime of interest, with a correlation length that is exponentially large in  $1/\Delta$.

\begin{figure}[t]
 \begin{center}
 \includegraphics[width=0.95\linewidth]{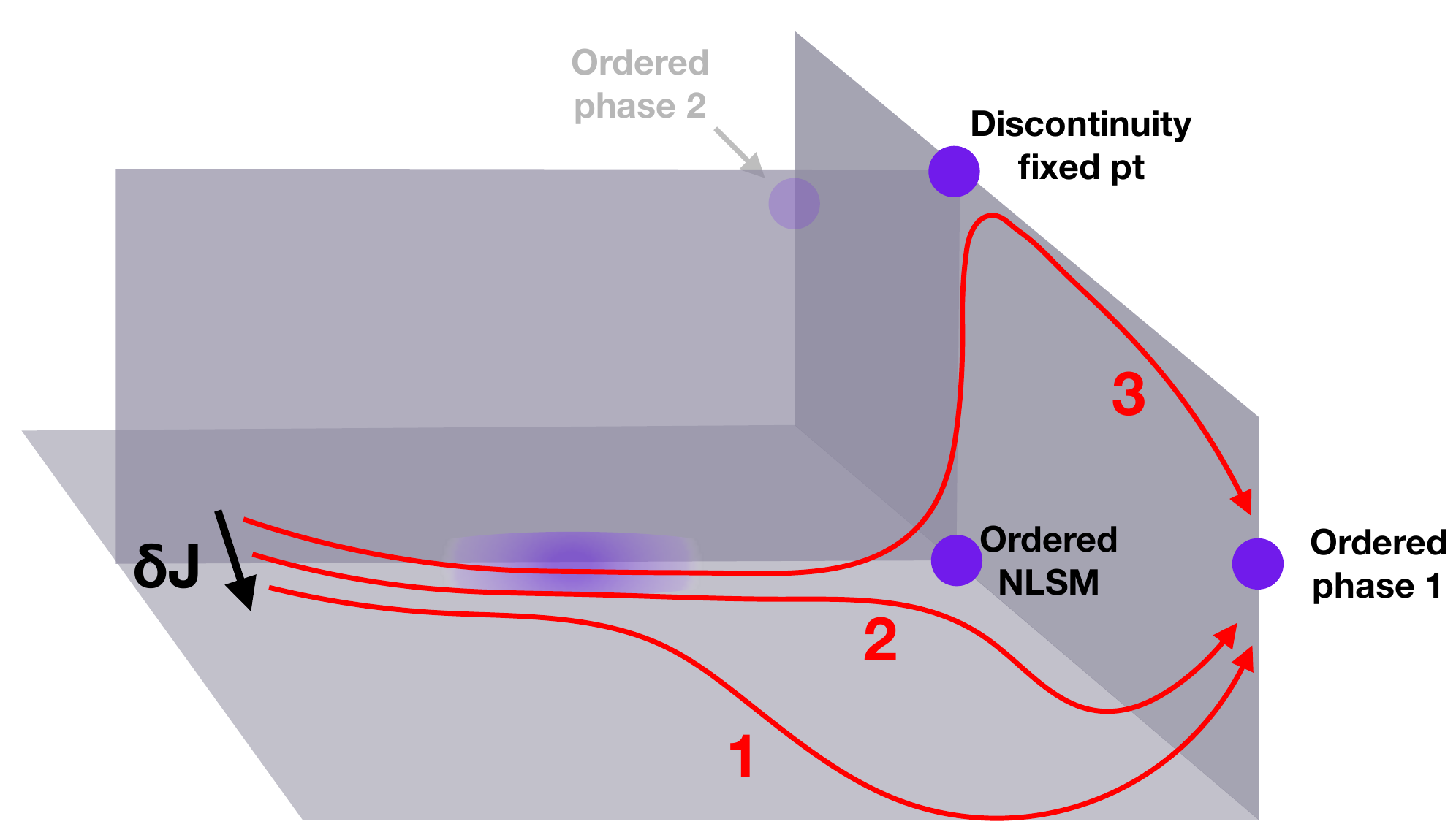}
 \end{center}
\caption{RG flows at small $\delta J = J-J_c$, showing three regimes 
in the idealized limit of an asymptotically weak first order transition (schematic).
Here both $g_R$ and $g_I$ are nonzero.
Flow line 1  comes close to the pseudocritical region before being driven away to the ordered phase by the relevant perturbation $g_R$.
Flow line 2,  closer to $J_c$, approaches the $O(4)$ long range ordered fixed point before being driven away.
Flow line 3, extremely close to $J_c$: in this regime we see the asymptotic `universal' properties of a standard first order transition.
}
\label{fig:deltajflows}
\end{figure}

Here we do not have a controlled $\Delta^2=0$ limit that preserves $O(4)$ symmetry, but to simplify the discussion,
let us imagine that such a limit could be found in an appropriately deformed field theory.\footnote{See  \cite{wang2017deconfined} for speculations about such deformations in the {5-component} case.}
This limit yields several well-defined regimes of RG flow,
giving the simplest illustration of the mechanism of interest.
The physical model we study is not in the controllable limit of parametrically small $\Delta$ 
(regardless of whether an appropriate deformation of the field theory can be found). 
Therefore, rather than giving quantitative information about the RG flows in the physical model,
 the following discussion motivates a conjecture for the qualitative features of those flows.
We  will see below that the expectations
from this conjecture appear
to be borne out in simulations.

As discussed in  \cite{wang2017deconfined,cardynauenbergscalapino},  for small $\Delta$ and $\lambda$ and in the vicinity of the flow line we have
\ba
\f{\dd \lambda}{\dd \ln L}  & = - \lambda^2 - a^2 \Delta^2 + \cdots \\
\f{\dd g_R}{\dd \ln L}  & = + (y_R + c' \lambda ) g_R+ \cdots, \\
\f{\dd g_I}{\dd \ln L}  & = -(y_I + c \lambda ) g_I + \cdots \label{eq:rgeqirr}
\end{align}
where $a$, $c$ and $c'$ are order one constants.
Here, in addition to $\lambda$, we have included the most relevant\footnote{More correctly, these couplings can be classified as relevant/irrelevant at the hypothetical $\Delta=0$ fixed point.}  coupling $g_R$ and the leading irrelevant one $g_I$ arising from $X^{(4)}$ (for simplicity we consider only one $g_I$).
We have extracted a sign so that  $y_R$ and $y_I$ are both positive, although $g_I$ is irrelevant.

First consider the case $g_R=g_I=0$.
Integrating the above equation shows that for small $\Delta$ the lengthscale required for $\lambda$ to flow from a positive order 1 value to a negative order 1 value scales as 
\be
\xi_{\text{order}} \sim e^{\pi/{a\Delta}}.
\ee
We make the \textit{assumption} that for large scales ${\ln L \gg \ln \xi_\text{order}}$ the flow is to the ordered fixed point of the sigma model in Eq.~\ref{eq:sigmamodel}.
This is consistent with the existence of a topological term in the sigma model since the topological term is irrelevant by power counting in the ordered phase.
This flow is indicated by the flow line marked {\bf A} in Fig.~\ref{fig:jcflows}.
The purple blur indicates the `slow' regime close to $\lambda = 0$.

Next consider the case $g_I\neq 0$, $g_R=0$. This is the flow line marked {\bf B} in Fig.~\ref{fig:jcflows}. 
Note that this  corresponds to the microscopic model precisely at $J_c$.
The perturbation $g_I$ is the leading effect of $O(4)$-breaking perturbations that are present in the microscopic model even at $J_c$.

The key point is that the $O(4)$--breaking perturbation $g_I$ decreases during the period of RG flow close to $\lambda=0$ (Eq.~\ref{eq:rgeqirr}). Since the coupling is effectively irrelevant ($y_I>0$), and since $\xi_\text{order}$ is exponentially large in $\Delta$, by the time $\lambda$ is negative and of order one, the irrelevant coupling $g_I$ has become exponentially small: $g_I^*\sim e^{-\pi y_I/a\Delta}$ \cite{wang2017deconfined}. 

Therefore the flow ${\bf B}$ comes very close to the ordered fixed point of the NLSM. 
However, the anisotropy encoded in $g_I$ is relevant at this ordered fixed point. In the next stage of the RG flow the flow line {\bf B} diverges from the ordered fixed point (Fig.~\ref{fig:jcflows}).\footnote{In the case of QED the leading symmetry-breaking terms in the sigma model involve two derivatives and four powers of the field and will have a much weaker effect \cite{tsmpaf06, wang2017deconfined}.}
This is similar to phenomena in symmetry-breaking transitions with dangerously irrelevant anisotropies \cite{lou2007emergence}.

Let us use the rescaled length coordinate $\tilde x = x/\xi_\text{order}$. Then an effective action in this regime is
\be
\label{eq:perturbedgoldstone}
\mathcal{S}= \int \dd^3 \tilde x \left[  \f{1}{2g_\text{eff}} 
\Big( \widetilde \partial {\bf n} \Big)^2 + {g_I^*} O_I \right],
\ee
where $O_I$ is an anisotropy in the sigma model with the appropriate symmetry.
Roughly speaking, the coupling $g_\text{eff}$ is of order one at the scale $\xi_\text{order}$, but grows rapidly under RG.
 Above, ${\bf n}^2=1$, but the lattice order parameters are related to this sigma model field  by a power of $\xi_\text{order}$, since the magnitude of the order parameter decreases roughly like a power of $L$ in the pseudocritical regime.

In fact there are two separate lengthscales associated with the $O(4)$ breaking perturbation $g_I^*$, depending on whether we consider the probability distribution of the ordered moment (the zero-wavevector mode of the field), which is essentially a four-component vector of fixed length living on the three-sphere, or the Goldstone modes. 
The probability distribution for the zero mode becomes asymmetric at a scale\footnote{Here we consider periodic boundary conditions, to avoid possible anisotropies associated with the boundary condition.}
\be
\xi_\text{zero mode} \sim \xi_\text{order} \times (g_I^*)^{-\f{1}{3}} \sim e^{ \f{\pi (1+y_I/3)}{a\Delta}}.
\ee
This is the lengthscale at which the contribution to the action (\ref{eq:perturbedgoldstone}) from the zero mode becomes of order 1, 
so that different orientations for the ordered moment are no longer approximately equally likely (this contribution is of order $g_I^*\times [L/\xi_\text{order}]^3$).
The correlation functions of the Goldstone modes cross over from power law to exponential on a different scale that is in principle longer,
\be
\xi_\text{Goldstone} \sim \xi_\text{order}\times (g_I^*)^{-\f{1}{2}} \sim e^{ \f{\pi (1+y_I/2)}{a\Delta}}.
\ee
(To see this note that in the rescaled coordinates $\tilde x$ the squared mass is simply proportional to $g_I^*$, since it is obtained   by expanding Eq.~\ref{eq:perturbedgoldstone} to quadratic order around one of the minima of the potential $O_I$.)

Finally, consider briefly the case where the microscopic model is perturbed away from the transition by ${\delta J = J-J_c}$.
There are three different regimes, depending on the size of $\delta J$, that are illustrated in Fig.~\ref{fig:deltajflows}.
The coupling $g_R$, which is initially small and proportional to $\delta J$, is growing at all stages of the RG flow. How small its initial value is will determine which fixed points in the $g_R=0$ plane the flow approaches before being driven away to one of the ordered phases at large (positive or negative) $g_R$, which here represent the N\'eel and VBS phases.

The above discussion is all in the idealized limit of small $\Delta$.
In this limit there is a clean separation of scales between $\xi_\text{order}$ and $\xi_\text{zero mode}$ corresponding to the regime `close' to the ordered fixed point of the symmetric sigma model  
(assuming we are sensitive to the probability distribution of the zero mode).

In the physical models, we do not expect this clean separation of scales, and we do not expect the detailed formulae in Eq.~\ref{eq:rgeqirr} to be accurate.
Roughly speaking, the lengthscales relevant to us below are on the scales of, say, $10$ to $100$ so the relevant RG times are only of order one, because of the logarithmic relationship between length and RG time. So we are certainly not in the above limit where the flow of the coupling becomes extremely slow. But conversely, the exponential dependence of irrelevant variables on RG time means that accurate $O(4)$ can still emerge on these scales.

We will give evidence that the RG flows in the model we study have the basic qualitative features described above: an initial regime in which approximate $O(4)$ arises, with $O(4)$ `order' at larger scales. (At present we cannot access large enough scales to see the subsequent flow \textit{away} from the symmetric sigma model at $J_c$, although this may be possible.)

Before continuing, we note one caveat. The advantage of the present scenario, as an explanation for the following numerical results, is that it does not assume any fine-tuning of the microscopic parameters of the model beyond tuning of $J$ to  $J_c$: the phenomenon is insensitive to the bare values of the irrelevant couplings $g_I$.
But in principle an alternative way to flow close to the ordered $O(4)$ fixed point would be to \textit{fine-tune} the bare parameters to be close to a tricritical \cite{qin2017duality,JianEmergent2017}  fixed point with \textit{two} unstable directions, {if} an appropriate such fixed point,  with $O(4)$ symmetry, happened to exist. Since this requires fine-tuning, we believe it is a less plausible explanation for our numerics than the scenario above.

\section{Model}
\label{sec:model}

The model we study is a simple modification  of the loop model in \cite{DCPscalingviolations}, so we summarise it briefly.
This modification was initially proposed by \cite{sernaunpublished} as a way to perturb the deconfined critical point to reach a first order transition. 
The model is an isotropic classical lattice model in three dimensions with cubic symmetry.

The relation to 2+1D deconfined criticality can be seen either from a transfer matrix construction or from an effective field theory \cite{DCPscalingviolations,PhaseTransitionsCPNSigmaModel}. The model has an exact $SO(3)$ spin symmetry which becomes manifest in a dual representation
(with a local N\'eel vector ${\vec N = (N_x, N_y, N_z)}$ that is defined on each link of the lattice model). Heuristically, the loops can be thought of as wordlines of spinons in a spin-1/2 magnet, with two colours associated with the two spin directions.

\begin{figure}[t]
 \begin{center}
 \includegraphics[width=0.9\linewidth]{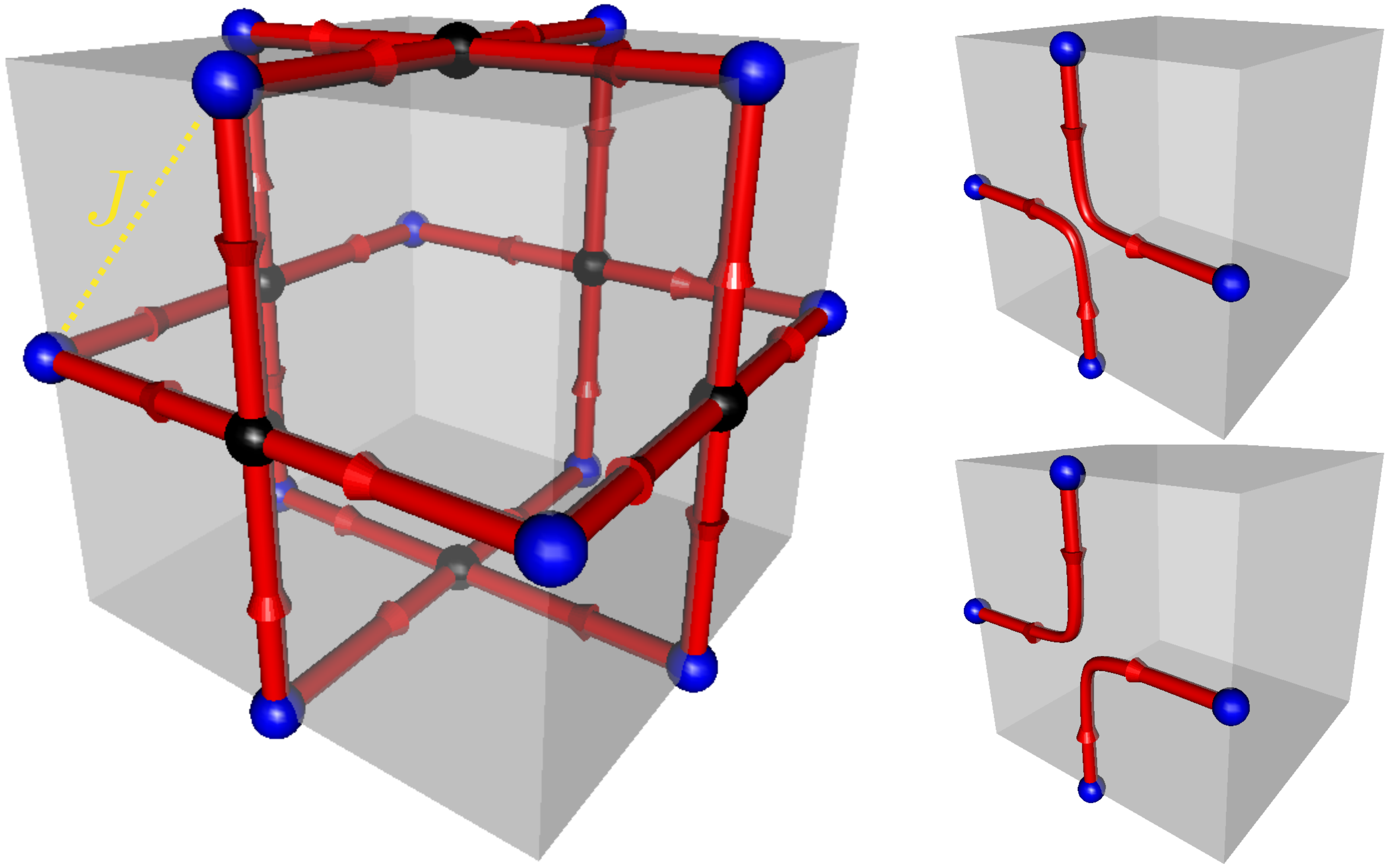}
 \end{center}
\caption{Left: Lattice structure of 3D L lattice. Nonintersecting loop configurations are obtained by choosing one of two possible pairings of links at each node $r$ (Right). There are interactions between nearest nodes on the A sublattice (shown in blue) of strength $J$ (indicated by dashed line). We set the interaction strength to zero for nodes on the $B$ sublattice (black).}
\label{fig:latticefigure}
\end{figure}

The partition function is an ensemble of completely-packed loop configurations on the `3D L lattice' \cite{cardy2010quantum} shown in Fig.~\ref{fig:latticefigure}.
(We use periodic boundary conditions, and measure lengths in units of the length of a link.)
At each node $r$ the loops can be connected in two possible ways, with the two options labelled by an Ising-like variable $\phi_r=\pm 1$; the sign convention for $\phi$ is defined below. 
The configuration $\{\phi_r\}$ determines the geometry of the loops. 
Each loop can also take one of two colours, which are summed over in the partition function:
\ba
Z & = \sum_{\{\phi\}} \sum_{\substack{\text{loop}\\\text{colours}}} e^{-E}.
\end{align}
Next we specify the energy $E$. The lattice is bipartite and each vertex $r$ is a member of either the $A$ or $B$ sublattice (blue and black in Fig.~\ref{fig:latticefigure}). There are interactions between nearest nodes on the \textit{same} sublattice. In \cite{DCPscalingviolations} these interactions were the same for the $A$ and $B$ lattices: we will refer to this here as the `symmetric' model.
 The modification here is to have a different interaction for the $A$ and $B$ sublattices:
\be\label{eq:energydefn}
E = - J_A \sum_{\< r, r'\> \in A} \phi_r\phi_{r'} 
- J_B \sum_{\< r, r'\> \in B}\phi_r \phi_{r'}.
\ee
The $\phi$ variables on the two sublattices make up the two components of the VBS operator (see below). As usual in deconfined criticality,  this operator also inserts a hedgehog  defect in the N\'eel order parameter (this can be verified by computing the Berry phase associated with a hedgehog \cite{DCPscalingviolations} in analogy to \cite{haldane, readsachdevberryphase}).

The model has a phase in which the average loop length diverges: this is the N\'eel phase, with long range $SO(3)$-breaking order. 
This phase occurs for small $J$. 
The N\'eel 2-point function is related to the probability that two distant links lie on the same loop, or (as is trivially related) to the probability that two distant links are the same colour. The loop representation allows efficient calculation of N\'eel correlators. In the following we will use $N_x$ to denote the uniform (spatially averaged) N\'eel order parameter. See the Supplementary Material of \cite{emergentso5} for more detail on how to measure correlation functions.

\begin{figure}[b]
 \begin{center}
 \includegraphics[width=0.9\linewidth]{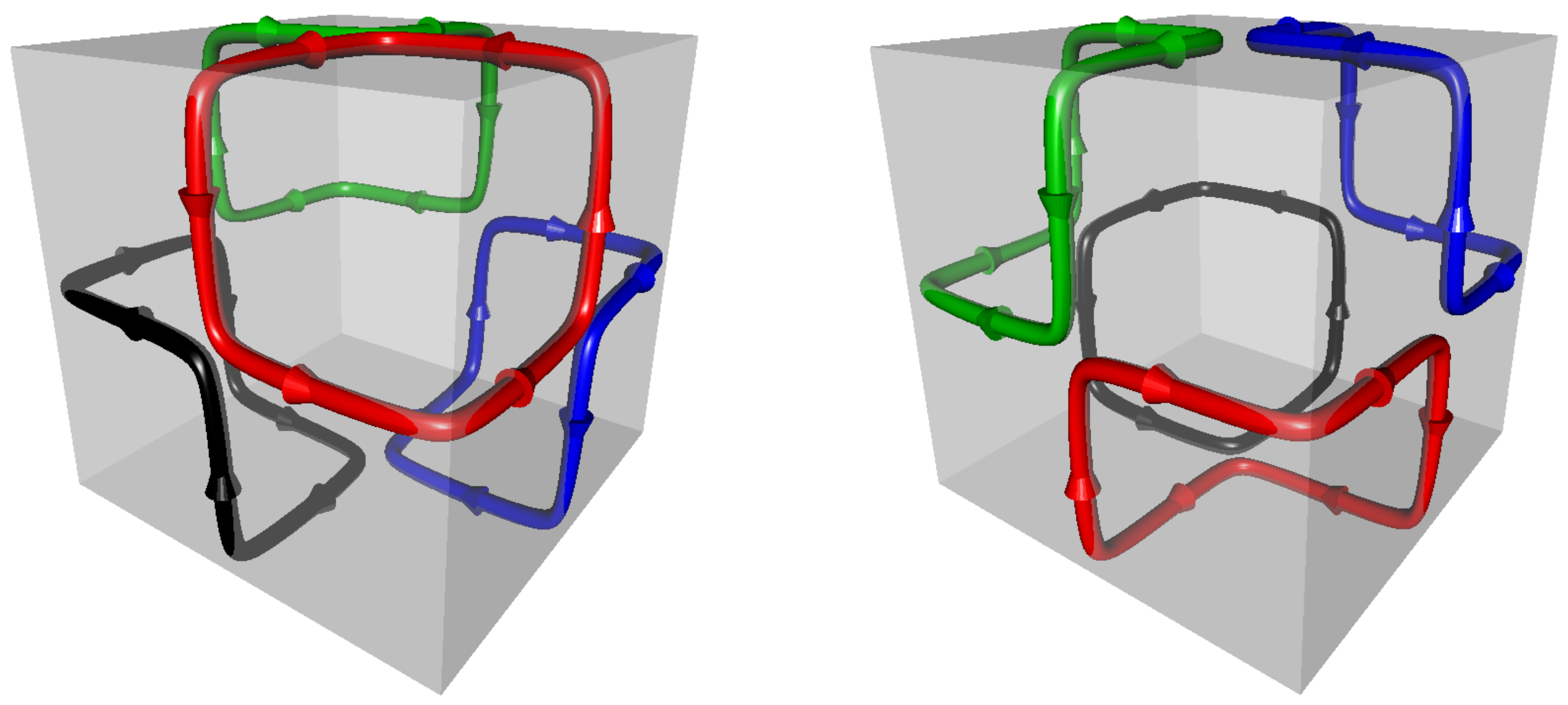}
 \end{center}
\caption{Two of the four equivalent configurations with minimal length loops (colours are only a guide to the eye). The configuration on the Left may be taken as the reference configuration to fix the sign convention for the $\phi$ variables.
In both configurations shown, the $A$ nodes have $\phi_x=1$. The $B$ nodes have $\phi_y=1$ in the Left figure and $\phi_y=-1$ in the Right figure, so these configurations are $(\phi_x, \phi_y)= (1,1)$ and $(1,-1)$ respectively.
We are interested in a phase transition at which $\varphi\equiv \phi_x$ orders while $\phi_y$ remains disordered.
The sign of $\phi_x$ can be viewed as picking out a sublattice of cubes for the loops to `resonate' around, see Fig.~\ref{fig:loopcubefig}.
}
\label{fig:signconvention}
\end{figure}

The model also permits VBS phases which break spatial symmetry and in which the loops are finite.
First consider the symmetric model, $J_A=J_B$.
At large enough $J_A=J_B$ the model is in a VBS phase that is four-fold degenerate.
This degeneracy can be understood from the extreme limit of large $J$
in which all the loops are of the minimal length (six links). 
There are precisely four such states. 
To fix the sign convention of $\phi_r$, we pick an arbitrary one of these four states as a reference state, and declare $\phi_r=1$ for all $r$ in this state (Fig.~\ref{fig:signconvention}).

The two components of the VBS order parameter are defined by $\phi_r$ for $r\in A$ and $r\in B$ respectively. 
The overall magnitude of the order parameter  is
\be\label{eq:phiuniform}
(\phi_x, \phi_y ) = \f{\sqrt 2}{\text{no. sites}} \lf \sum_{r\in A} \phi_r, \sum_{r\in B} \phi_r \ri.
\ee
In the four extreme VBS states above (which are related by lattice symmetry) we have
\be\label{eq:orderedstates}
(\phi_x, \phi_y ) = \f{1}{\sqrt 2} (\pm 1, \pm 1).
\ee
The component $N_z$ of the N\'eel order parameter may be mapped to a variable $\chi_\ell$, living in the link $\ell$, 
whose value is $\pm 1$ depending on whether the link is red or blue \cite{emergentso5}. The spatially averaged order parameter is then
${N_z = \sum_{\ell} \chi_\ell / (\text{no. links})}$.

\begin{figure}[t]
 \begin{center}
 \includegraphics[width=0.85\linewidth]{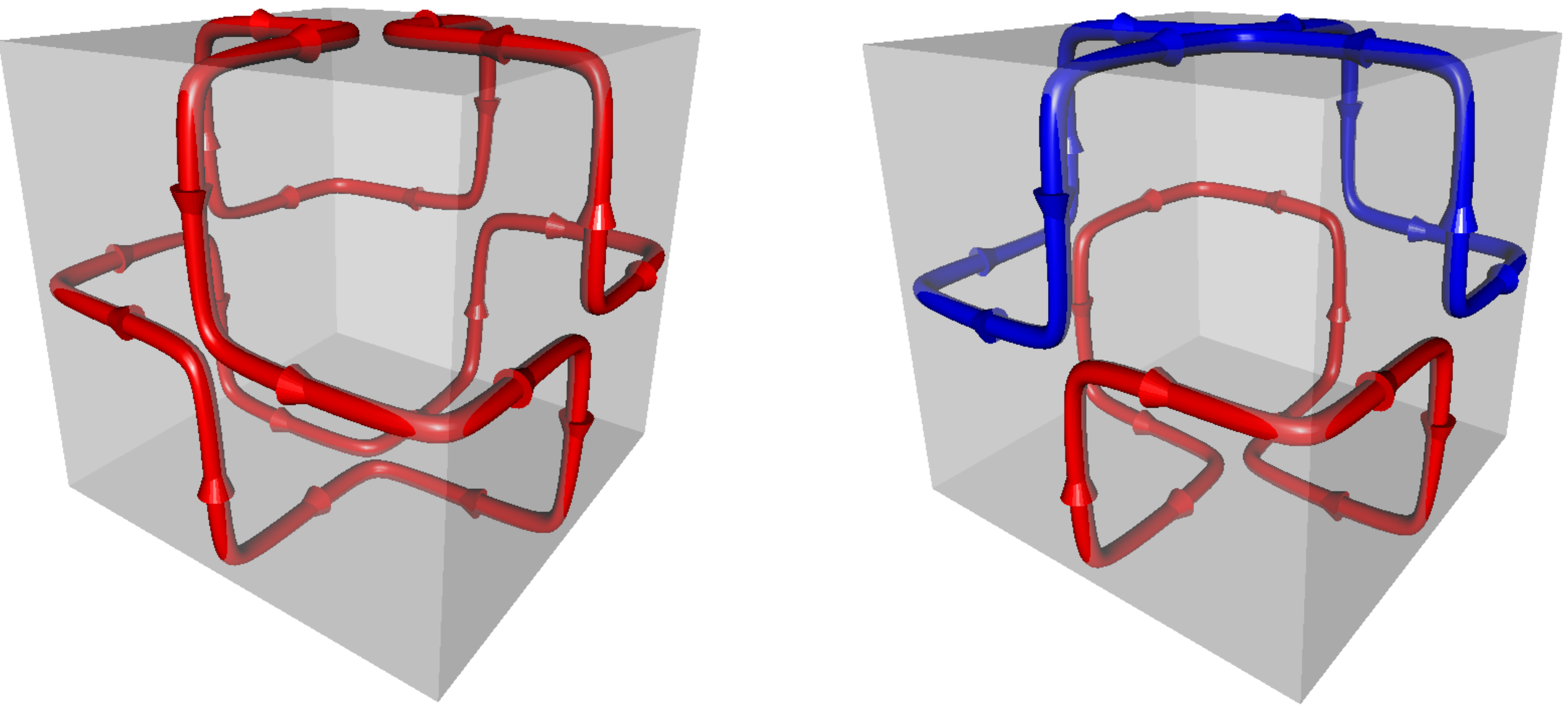}
 \end{center}
\caption{
When $\phi_x$ is ordered, but $\phi_y$ is not, this picks out a sublattice of cubes. The  figures show the case where $\phi_x=1$ and  $\phi_y$ is arbitrary.
In this extreme limit the loops live on the cubes with centres at $2(a,b,c)$ with ${a+b+c=\text{even}}$, such as the cube shown centred at the origin. For $\phi_x=-1$, the loops live on cubes with centres at $2(a,b,c)$ with ${a+b+c=\text{odd}}$ (the length of one link is taken to be 1).
}
\label{fig:loopcubefig}
\end{figure}

In the symmetric model the transition from N\'eel to VBS is at $J_A=J_B=0.088501(3)$.
In this work we will fix $J_B$ to \textit{zero}, in order to be far away from the symmetric critical point (avoiding crossover effects), and we will vary $J_A$ to access the phase transition between the N\'eel and VBS phases:
\ba
\label{eq:jajb}
J_A & \equiv J & 
J_B & = 0\,.
\end{align}
We take $J>0$. Therefore $\phi_x$ is more strongly coupled than $\phi_y$.
As $J$ is increased, there is a phase transition at which N\'eel order disappears, and $\phi_x$  orders. At this N\'eel-VBS phase transition, $\phi_y$ remains disordered (massive). 
This is the transition that we will study.
The sign of the order parameter $\<\phi_x\>$ can be viewed as picking out one of two sublattices of cubes for the loops to `resonate' around, see Fig.~\ref{fig:loopcubefig} for a cartoon.

If, starting in the VBS phase with $\<\phi_x\>\neq 0$, we then increase $J_B$ sufficiently, we encounter another phase transition at which $\phi_y$ also orders. This is a conventional Ising transition which we will not be interested in.
The full phase diagram for the model, close to the symmetric critical point, is shown in Fig.~\ref{fig:phasediagram1}.

In Appendix~\ref{sec:twovbs} we show data for both components of $\phi$ at the transition of the model with  $J_B=0$. This confirms that $\phi_y$ is a massive field with a short correlation length, and that for these parameters the model is `far' from the symmetric critical point in the model with ${J_A=J_B}$: i.e. there is a strong asymmetry between $\phi_x$ and $\phi_y$ even at short scales.
It is important to check this, because the critical point in the  \textit{symmetric} model is known to  have a very accurate $SO(5)$ symmetry. If we studied a very weakly asymmetric model that was too close to this symmetric critical point, there would be an intermediate range of scales with approximate $SO(5)$, and $O(4)$ would be a trivial consequence of this higher symmetry. This is not the case here.

The order parameters that are involved in the phase transition of interest  (indicated by the pink arrow in Fig.~\ref{fig:phasediagram1}) are $\phi_x$ and $\vec N$. To minimize subscripts we will write
\be
 \varphi \equiv \phi_x.
\ee
The order parameters can be arranged in a superspin $\vec n$, Eq.~\ref{eq:superspin}. 
Since the normalization of the lattice order parameters is arbitrary, $\varphi$ must be rescaled by a fixed but nonuniversal constant in order for $\vec n$ to have a chance of being invariant under $O(4)$ rotations.

Note that the relevant symmetry is $O(4)$ rather than $SO(4)$: improper rotations of ${\bf n} = (N_x, N_y, N_z, \varphi)$ are allowed. A translation by two link lengths parallel to an axis is a microscopic example of an improper $O(4)$ rotation, as it changes the sign of $\varphi$ but not of $\vec N$. [This translation also changes the sign of $\phi_y$, so in the symmetric model it is however a \textit{proper} rotation within $SO(5)$.]

\begin{figure}[t] 
 \begin{center}
 \includegraphics[width=0.7\linewidth]{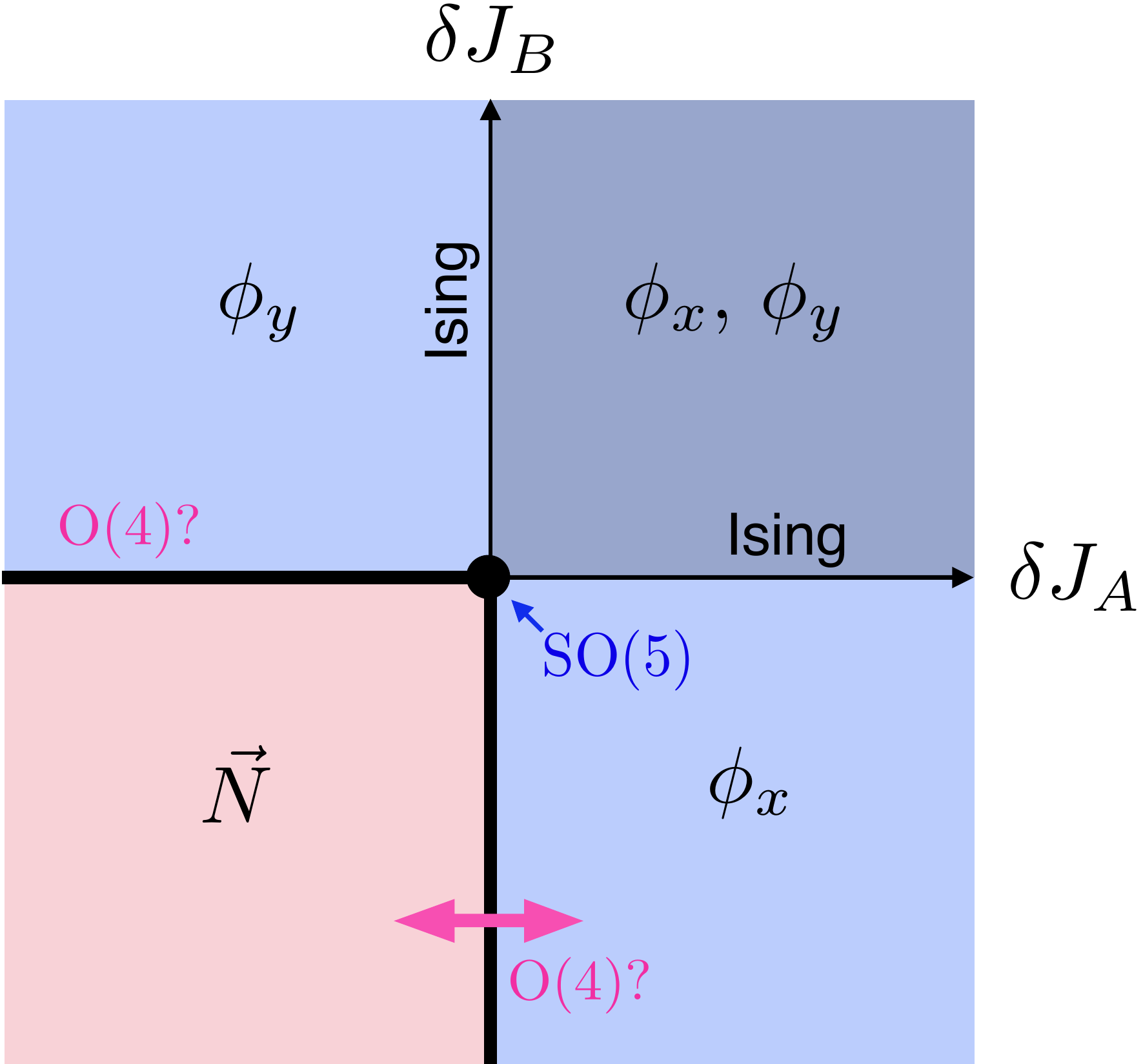} 
 \end{center}
\caption{
Schematic phase diagram of the model in Eq.~\ref{eq:energydefn} in the vicinity of the symmetric critical point at ${J_A=J_B\simeq 0.0885}$. The N\'eel phase and the three VBS phases are labelled by the order parameter(s) with a nonzero expectation value in the phase. We study the transition indicated by the pink arrow. However we study this transition relatively \textit{far} from the symmetric critical point, in order to avoid crossover effects.
}
\label{fig:phasediagram1}
\end{figure}

Before turning to numerical results, we  briefly clarify our notation for the VBS order parameter, and  contrast the present model with a  square lattice quantum magnet with rectangular anisotropy in the couplings \cite{tsmpaf06, metlitski2017intrinsic}.

Our  convention for $\vec{\phi}$ is such that  the VBS-ordered states in the symmetric model ($J_A=J_B>J_\text{critical}$)  are at $\vec \phi\propto (\pm 1, \pm 1)$.
If we temporarily redefine the order parameter by ${\vec \phi' = \f{1}{\sqrt 2} ( \phi_x + \phi_y, \phi_x - \phi_y)}$, the ordered states are instead at $\vec \phi' \propto (\pm 1, 0)$, $(0, \pm 1)$.
This now matches the usual convention for columnar VBS ordered states in square lattice quantum magnets. 

In this language, making  $J_A$ slightly different from $J_B$, as above, introduces the term
${\phi_x^2 - \phi_y^2 \propto \phi_x' \phi_y'}$, which preserves symmetries under reflections across the lines ${\phi'_x = \pm \phi'_y}$. 
If we start in the VBS-ordered phase of the symmetric model and turn this perturbation on weakly, these symmetries ensure that there are still four degenerate ground states: this corresponds to the upper right quadrant of Fig.~\ref{fig:phasediagram1}.

\begin{figure}[t]
 \begin{center}
 \includegraphics[width=0.7\linewidth]{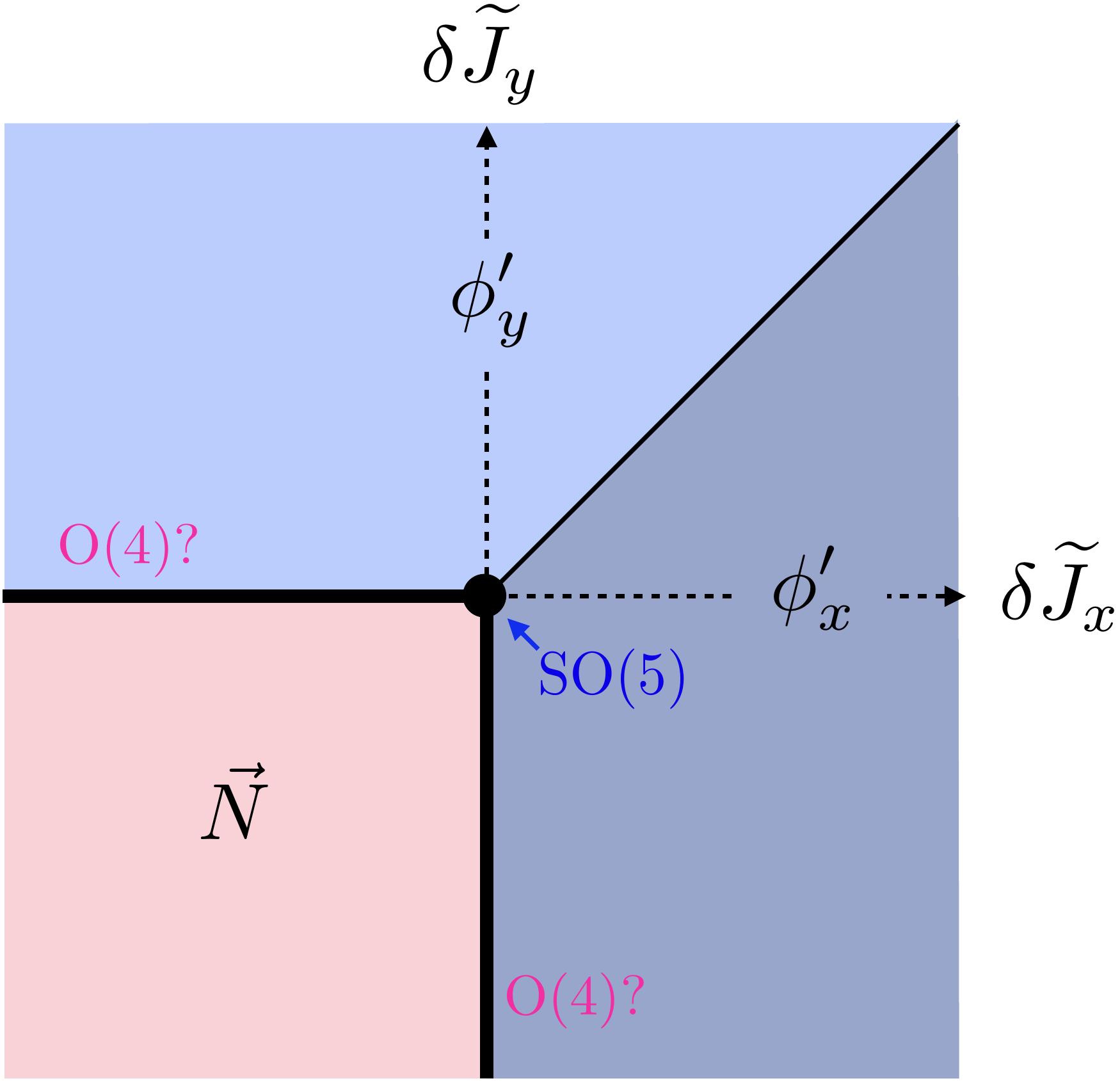} 
 \end{center}
\caption{
We contrast Fig.~\ref{fig:phasediagram1} with the phase diagram expected when a perturbation ${\delta\tilde J_x - \delta\tilde J_y}$,  favouring two out of four of the VBS minima, is added to the critical point in the symmetric model (see text for notation). 
This perturbation is analogous to rectangular anisotropy in a square lattice quantum magnet at the N\'eel to columnar VBS transition, see text.
}
\label{fig:phasediagram2}
\end{figure}

This should be contrasted with rectangular anisotropy in the square lattice magnet, which is analogous to a different perturbation  of the Lagrangian, ${(\phi'_x)^2 - (\phi'_y)^2}$, which instead preserves reflection symmetries across the lines $\phi'_x = 0$ and $\phi'_y=0$.\footnote{While this perturbation is related to the previous one by a $\pi/4$ rotation of the order parameter,  higher order terms in $\phi$ that break the $U(1)$ symmetry in the ordered phase are not invariant under this rotation.}
In the VBS ordered phase, turning this perturbation on immediately reduces the ground state degeneracy from four to two.
The phase diagram for a model with rectangular anisotropy,  perturbed away from the symmetric critical point by ${\delta \mathcal{L} = \delta \widetilde J_x (\phi'_x)^2 + \delta \widetilde J_y (\phi'_y)^2}$, is shown in Fig.~\ref{fig:phasediagram2}. The diagonal line is the four-fold degenerate VBS-ordered phase of the symmetric model.

Despite the different topologies of Figs.~\ref{fig:phasediagram1},~\ref{fig:phasediagram2}, 
 if there is indeed a quasiuniversal $O(4)$ regime that is accessible via an RG flow from the $SO(5)$ point,
 then we would expect to be able to access it in both models (at least close enough to the symmetric critical point) on the phase transition lines marked  `$O(4)$?' in Figs.~\ref{fig:phasediagram1},~\ref{fig:phasediagram2}.
 
As discussed above, simulations of the phase transition close to the symmetric point would be complicated by additional crossover effects, so here we study a point on the phase transition line that is relatively far from the line $J_A=J_B$.

\section{Numerical results}
\label{sec:numerics}

\begin{figure}[b]
 \begin{center}
 \includegraphics[width=0.95\linewidth]{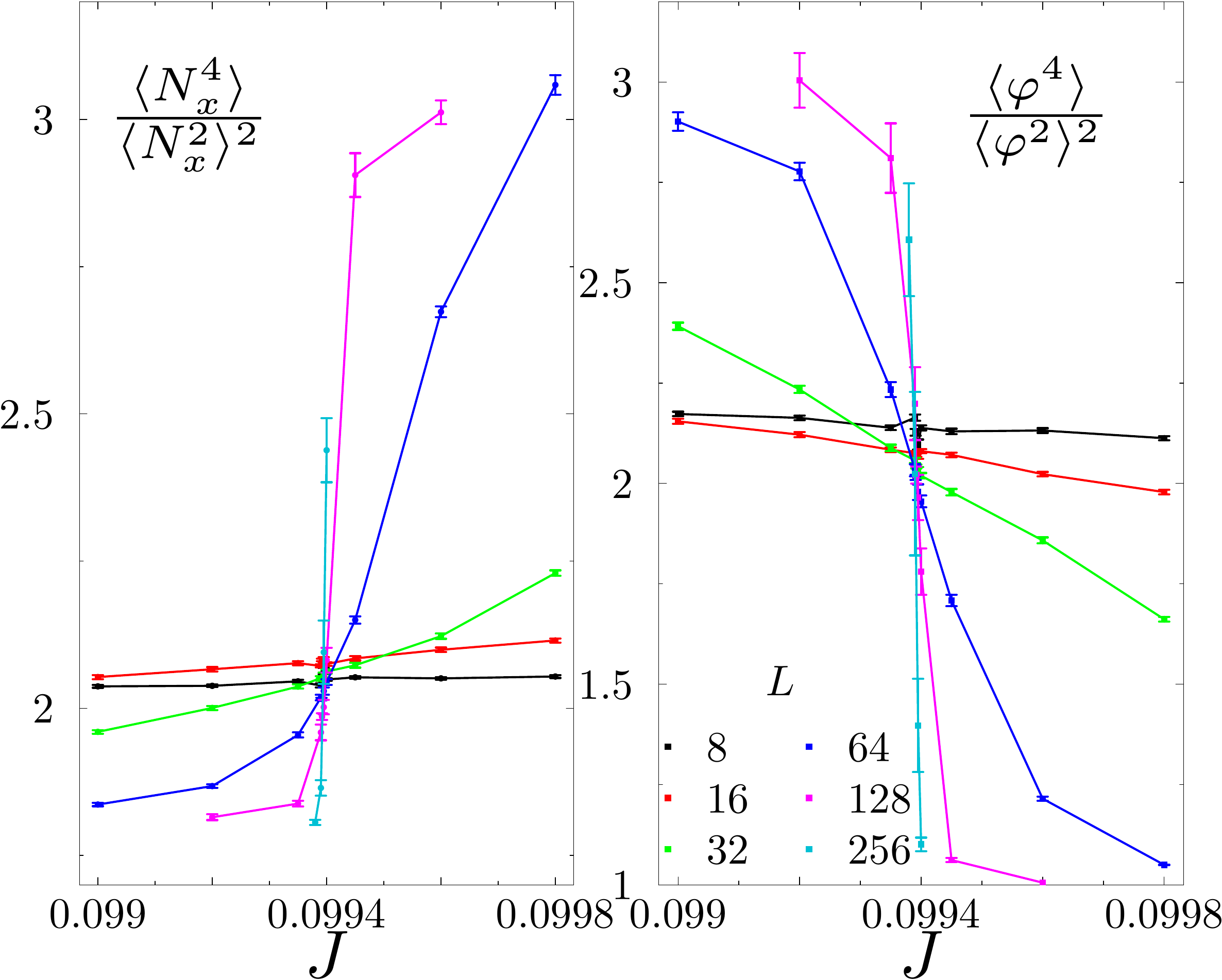}
 \end{center}
\caption{ Binder cumulants $\langle N_{x}^{4}\rangle/\langle N_{x}^{2}\rangle^{2}$ (left panel) and $\langle \varphi^{4}\rangle/\langle \varphi^{2}\rangle^{2}$ (right panel)  shown as a function of the coupling $J$, for several system sizes. 
}
\label{fig:binder}
\end{figure}

In Fig.~\ref{fig:binder} we show the Binder cumulants for the two order parameters close to the critical coupling,
\be
\label{eq:couplingestimate}
J_c = 0.0993916(10)
\ee
(our method for estimating $J_c$ is described below).
At first glance these data for the Binder cumulants appear to show a continuous transition between phases with N\'eel and VBS order.
Note the absence, on these scales, of the diverging positive peaks 
that would be expected for a conventional first order transition. We will argue that the transition is ultimately first order, but has a broad regime of lengthscales where it resembles an $O(4)$ spin flop transition.
On this range of lengthscales, the N\'eel and VBS states are effectively related by a continuous rotational symmetry.
This is quite different from a conventional first order transition, where, at $J_c$, the two competing phases are not even related by a discrete symmetry, and have (in the classical case) different energy and entropy.

We argue in three steps: first for the first-order nature of the transition, then for the presence of $O(4)$ symmetry, then for the presence of effective $O(4)$ long range order.

\subsection{Evidence for first order nature of transition}
\label{subsec:firstorder}

\begin{figure}[t]
 \begin{center}
 \includegraphics[width=0.95\linewidth]{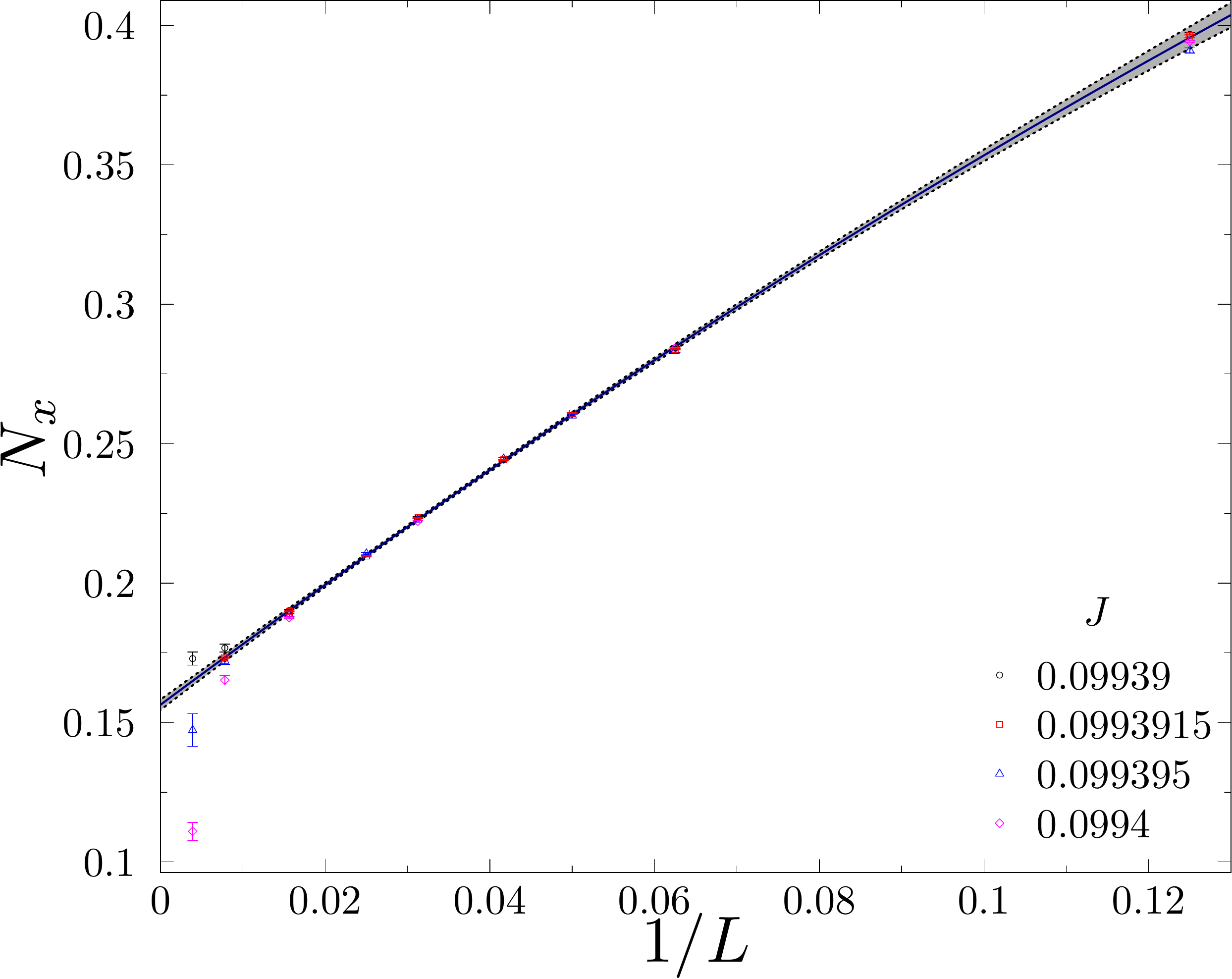}
 \includegraphics[width=0.95\linewidth]{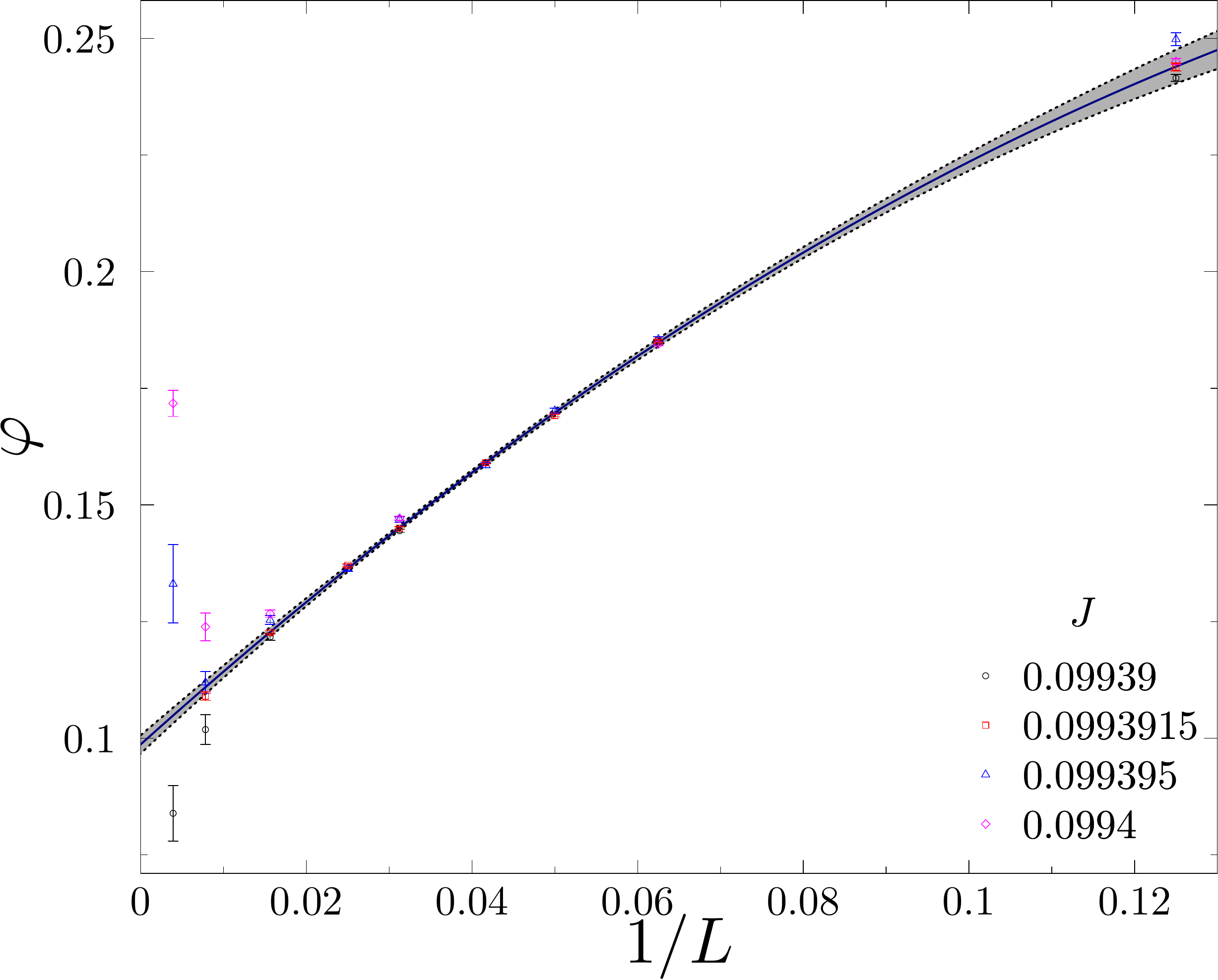}
 \end{center}
\caption{Order parameters $N_{x}$ and $\varphi$ as a function of $1/L$ for several values of $J$ close to $J_{\rm c}$ . Data points at $J=0.0993915$ are fitted to ${a+b/L+c/L^{2}}$, shown as a continuous line together with the shaded area corresponding to the statistical 95\% confidence interval.
}
\label{fig:Oatjc}
\end{figure}

We first give evidence that the transition is first order rather than continuous.
Since the transition is certainly not strongly first order, we cannot access the very large system sizes $L$ required to see conventional first order signatures such as double-peaked histograms for the energy.
Instead we examine the size-dependence of the order parameters at $J_c$. 
We give evidence that as $L\rightarrow\infty$ both order parameters extrapolate to a nonzero value, indicating coexistence between the two phases at $J_c$, i.e. a first order transition.

The two panels in Fig.~\ref{fig:Oatjc} show the N\'eel and VBS order parameters, $N_{x}\equiv \sqrt{\<N_x^2\>}$ and $\varphi \equiv \sqrt{\<\varphi^{2}\>}$,
 as a function of system size at four selected values of $J$ close to $J_{c}$. (Here the operator in the expectation value, e.g. $N_x$, is averaged over the  system volume.)
 The continuous lines 
are fits to the form $a+b/L+c/L^{2}$  for the $J$ value closest to $J_c$
(shaded areas are 95\% confidence intervals for these fits). Extrapolated values at $L=\infty$ are:
\ba\label{eq:extrapolatedorderparams}
N_{x}&=0.1562 (14), &
\varphi &=0.0987(14).
\end{align}
Fits to similar forms, including $a+b L^{-c}$, give compatible results, while putting the data on a log-log plot (Appendix~\ref{sec:twovbs}) shows that a power-law fit with $a=0$ is not good.
Note that the numerical values in Eq.~\ref{eq:extrapolatedorderparams} are not expected to be the same for both order parameters, as they depend on the arbitrary normalizations of the lattice operators.

\subsection{Evidence for approximate $O(4)$ symmetry}
\label{subsec:moments}

If $O(4)$ symmetry emerges it fixes relations between various moments of the order parameters at $J_c$. 
We will give evidence for approximate symmetry under a subgroup of $O(4)$, whith is the $U(1)$ acting on $(N_x, \varphi)$. 
(The $SO(3)$ acting on $\vec N$ is an exact microscopic symmetry.)
Given microscopic symmetries, an additional emergent symmetry rotating $N_x$ into $\varphi$ is sufficient\footnote{See appendices of \cite{sreejith2018emergent} for a discussion in the $SO(5)$ case.} to generate the full emergent internal $O(4)$ symmetry.

\begin{figure}[t]
 \begin{center}
 \includegraphics[width=0.95\linewidth]{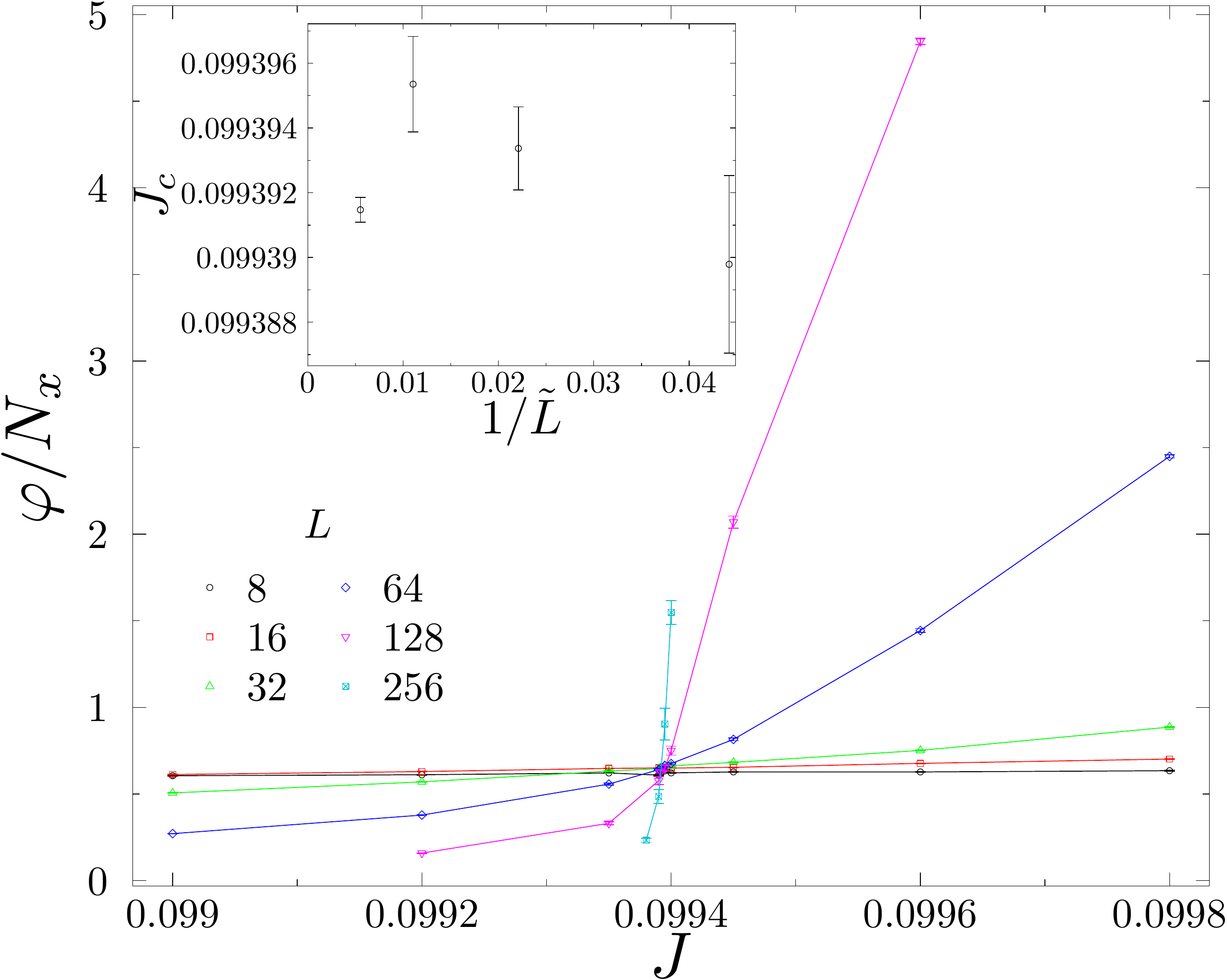}
 \end{center}
\caption{ Main panel: curves of the ratio $\sqrt{\<\varphi^2\>/\<N_x^2\>}$ as function of the coupling $J$, showing a clear crossing close to $J\approx0.0993915$. Inset: values of $J$ at the crossing points between two consecutive system sizes, $L_{1}$ and $L_{2}$, as a function of the inverse of the geometric mean $\tilde L =\sqrt{L_{1}L_{2}}$.
}
\label{fig:phioverN}
\end{figure}

First we examine the ratio between the order parameters, ${\varphi/N_{x} \equiv \sqrt{\<\varphi^2\>/\<N_x^2\>}}$, in Fig.~\ref{fig:phioverN}.
In the presence of emergent symmetry, this ratio should be $L$-independent at $J_c$ (see e.g. \cite{emergentso5,sreejith2018emergent}), yielding a crossing of curves for different $L$.  Fig.~\ref{fig:phioverN} indeed shows a well-defined crossing if the  smallest size, $L=8$, is excluded, despite the fact that the order parameters  are individually varying strongly with $L$.

The $J$-values of the crossings between consecutive $L$s are shown in the  inset to Fig.~\ref{fig:phioverN}. A weighted average of these crossings yields the estimate $J_{c} = 0.0993918(7)$ for the transition point.

\begin{figure}[t]
 \begin{center}
 \includegraphics[width=0.95\linewidth]{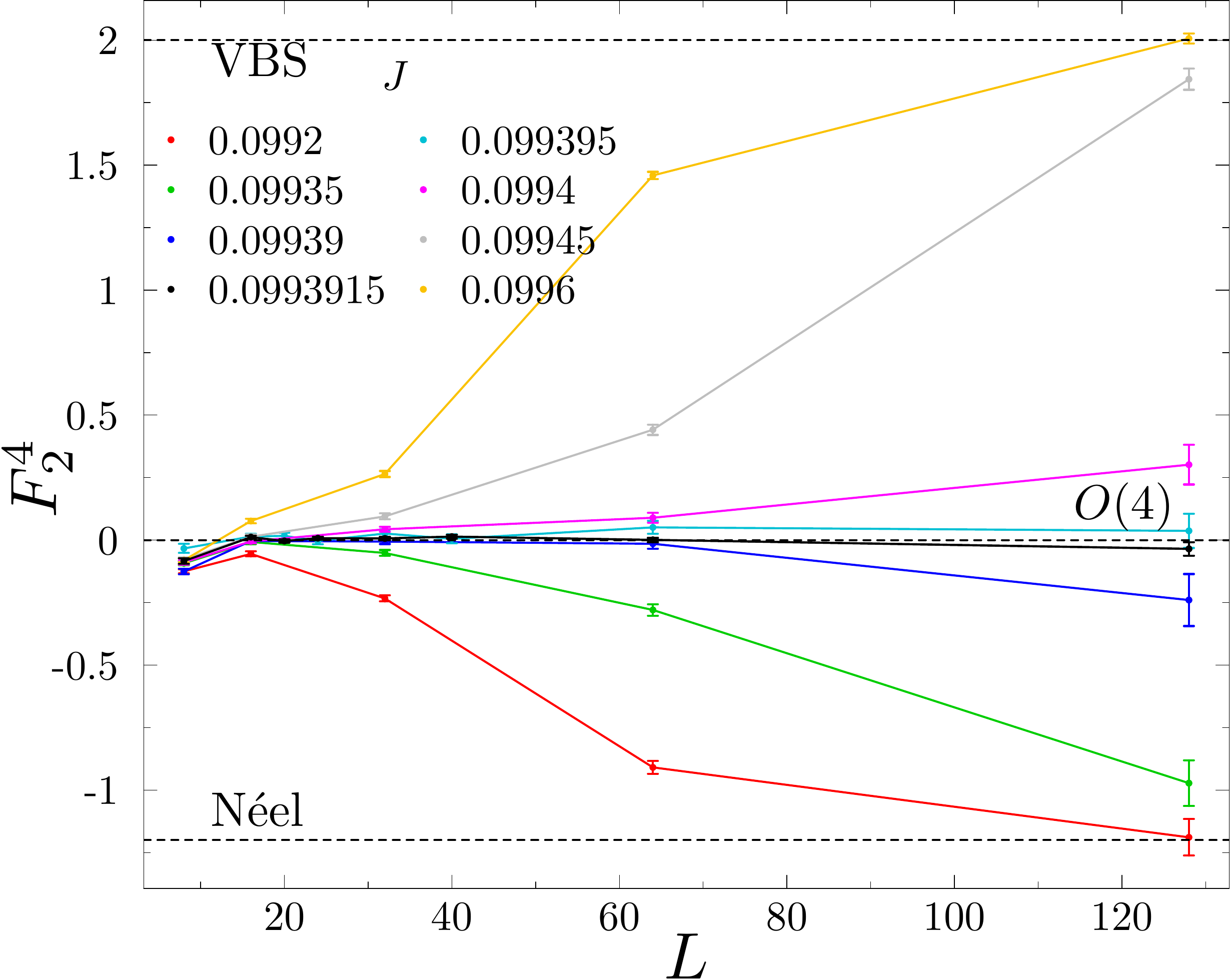}
 \includegraphics[width=0.95\linewidth]{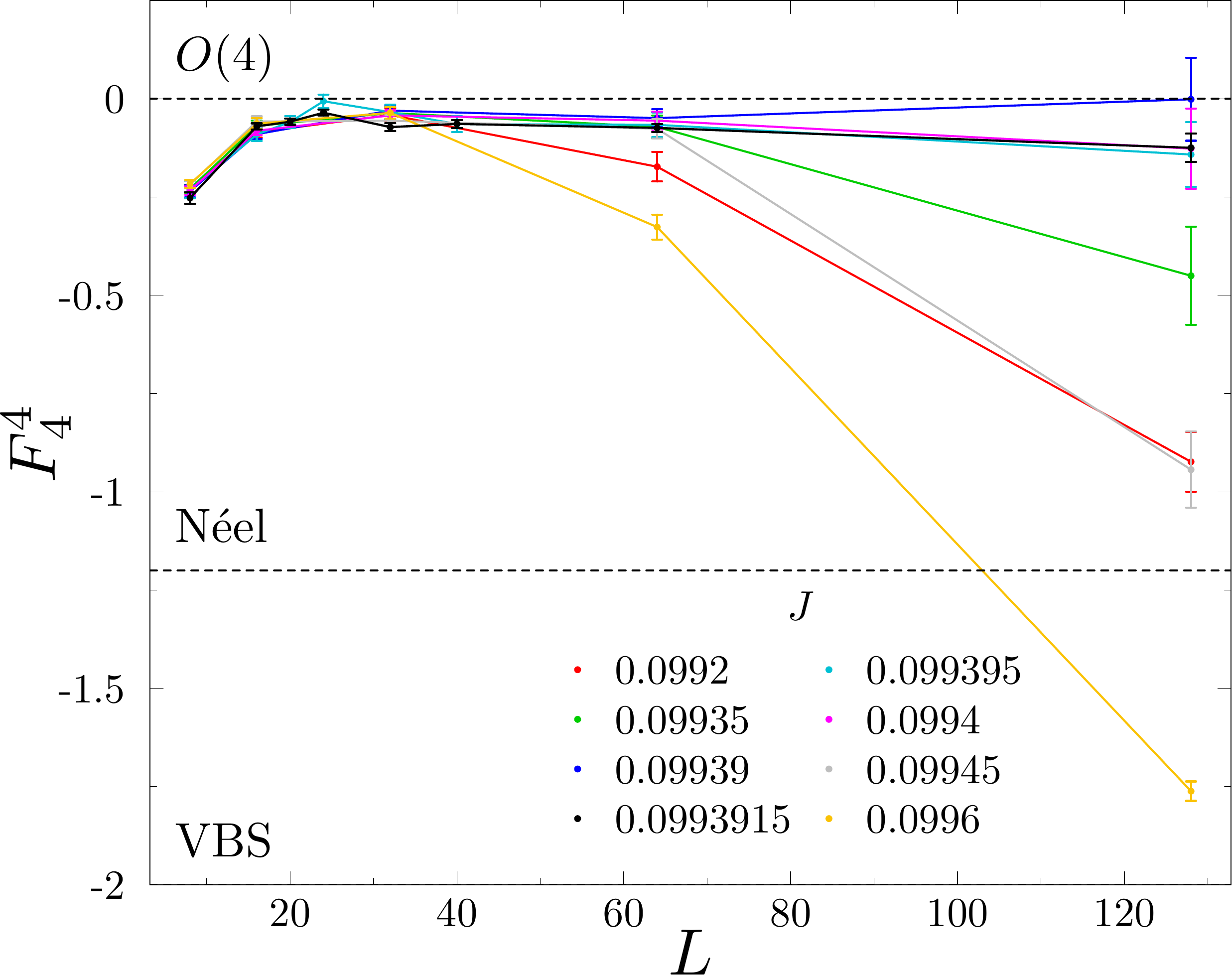}
 \end{center}
\caption{Moments $F_{2}^{4}$ and $F_{4}^{4}$ as a function of the system size for several values of $J$. Dashed lines correspond to expected values in the N\'eel phase, at an $O(4)$-symmetric point, and in the VBS ($Z_{2}$) phase.
}
\label{fig:F24vsL}
\end{figure}

\begin{figure}[t]
 \begin{center}
 \includegraphics[width=0.95\linewidth]{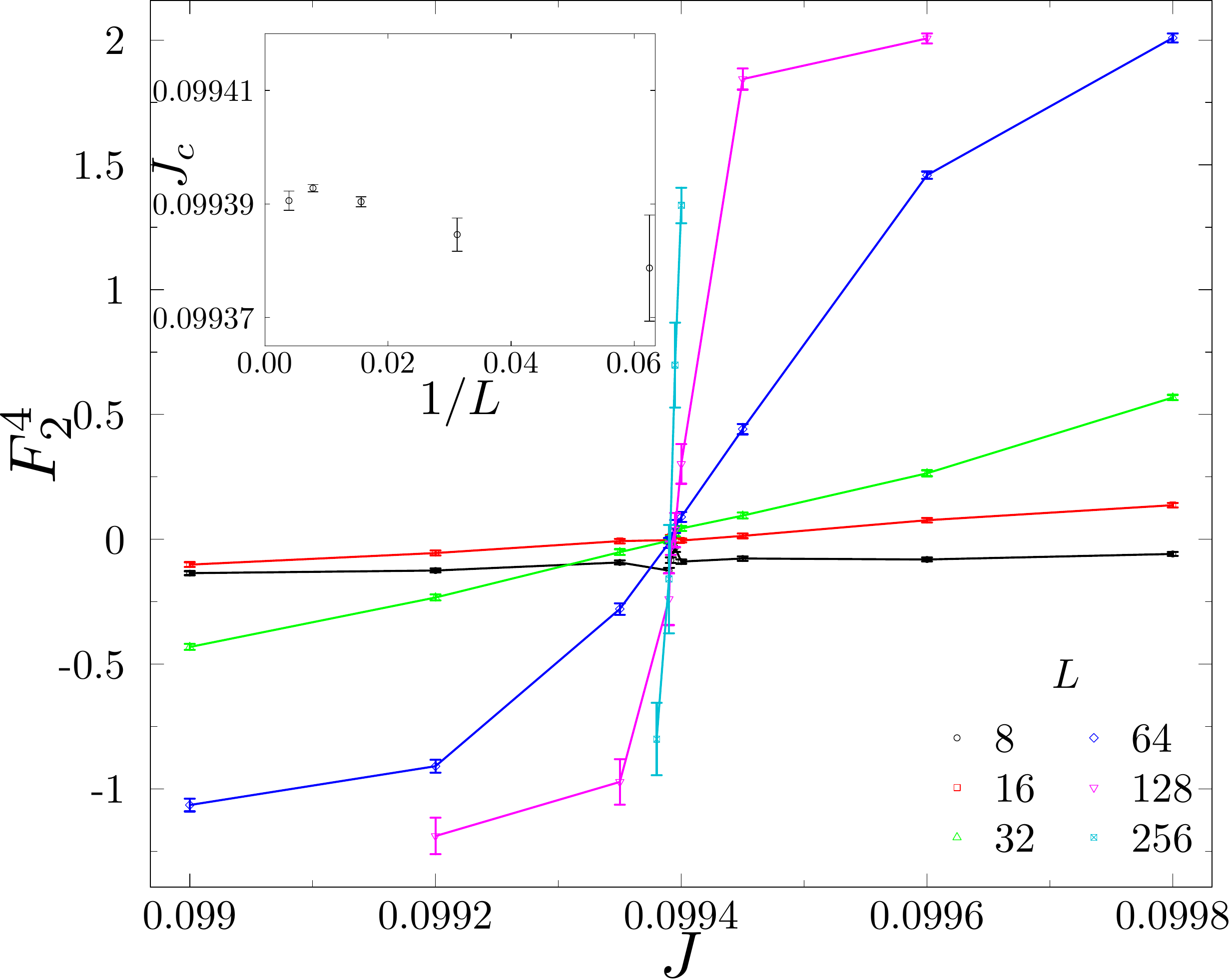}
  \includegraphics[width=0.95\linewidth]{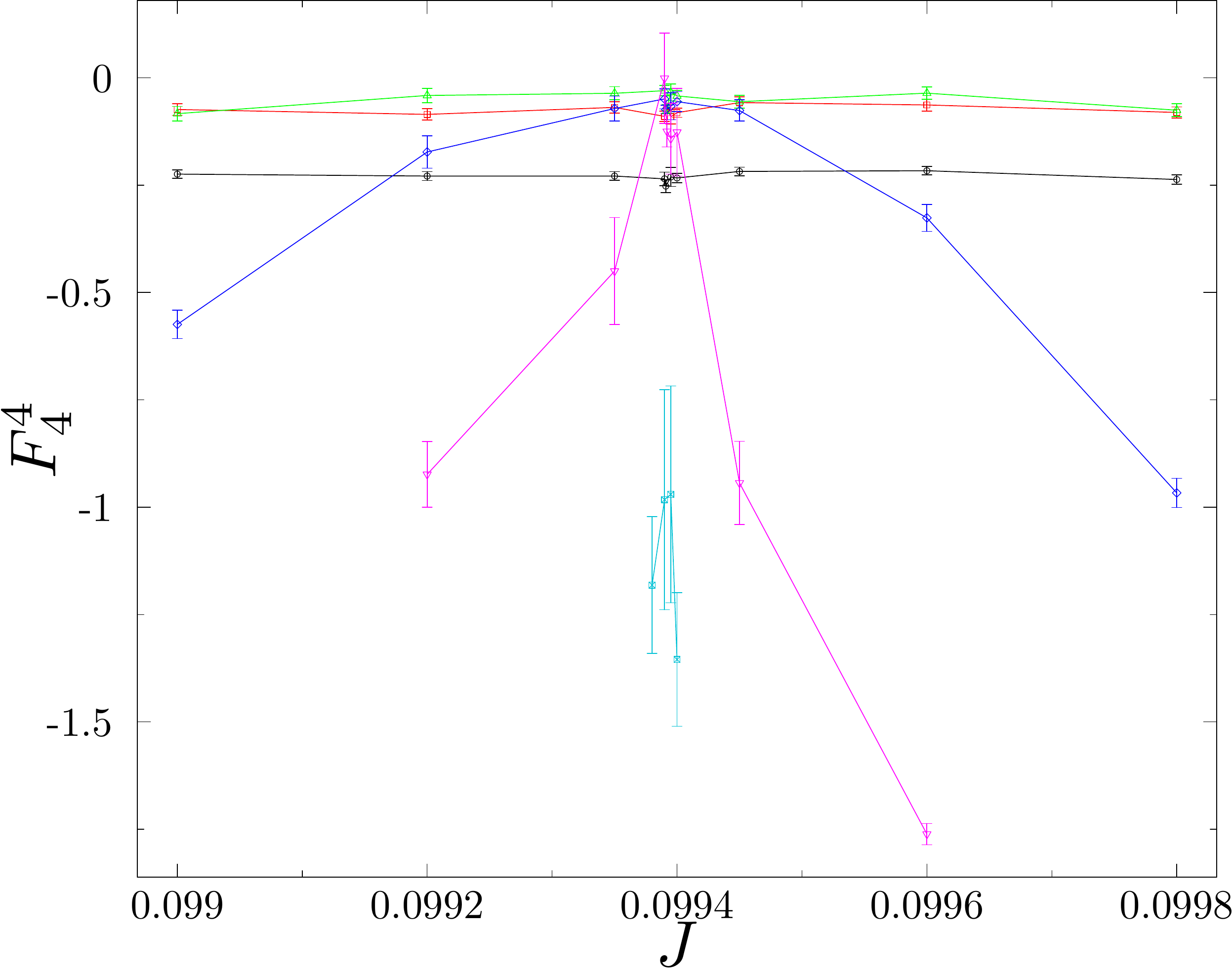}

 \end{center}
\caption{Moments $F_{2}^{4}$ and $F_{4}^{4}$ as a function of $J$. 
Note that in the second panel, the peak in $F_4^4$ for $L=256$ is probably hidden because its width is smaller than the separation between the data points.
Inset: values of $J$ at which $F_{2}^{4}$ is equal to 0, shown versus $1/L$. The weighted average gives an estimate of $J_c$.
}
\label{fig:F24vsJ}
\end{figure}

Next we examine some moments of the joint probability distribution for $N_x$ and $\varphi$ \cite{emergentso5,sreejith2018emergent}.
Let ${(N_x, \varphi) = r (\cos\theta, \sin\theta)}$, and let
 $F^{a}_{\ell}=\< r^{a}\cos( \ell \theta)\>$. This should \textit{vanish} for $\ell>0$ in the presence of $U(1)$ symmetry relating these two order parameter components.
 
In Figs.~\ref{fig:F24vsL}~and~\ref{fig:F24vsJ} we show
\be
F_{2}^{4} = \<\tilde N_{x}^{4} - \tilde \varphi^{4}\>,\quad
F_{4}^{4}= \< \tilde N_{x}^4 - 6 \tilde N_{x}^2 \tilde \varphi^2 + \tilde \varphi^4 \>.
\ee
Here $\tilde N_{x} = N_{x}/\sqrt{\<N_{x}^{2}\>}$ and $\tilde \varphi = \varphi/\sqrt{\<\varphi^{2}\>}$ are the order parameters normalized to have unit variance.

In Fig.~\ref{fig:F24vsL} the quantities $F_2^4$ and $F_4^4$ are shown as a function of system size for various $J$ values close to $J_c$.
Strikingly, at ${J=0.0993915}$ (the value closest to $J_c$)
$F_{2}^{4}$ is zero to within  errors for system sizes $16\le L\le 128$.
$F_{4}^{4}$ approaches zero quite closely, consistent with approximate $O(4)$. There  appears to be a small but  measurable difference from zero over this size range, indicating that $O(4)$ is not perfect.

$F_2^4$ and $F_4^4$ are also plotted as a function of $J$ in 
Fig.~\ref{fig:F24vsJ}.
The $J$ values of the crossings of $F_{2}^{4}$ with zero,
which are shown in the inset of Fig.~\ref{fig:F24vsJ},
can be used to obtain a different estimate of the transition point, which is ${J_{c} = 0.0993916(10)}$.
This is the estimate in Eq.~\ref{eq:couplingestimate}.

The curves for $F_4^4$ in the lower panel of Fig.~\ref{fig:F24vsJ}  feature a peak at $J_c$ with a maximum close to zero. This peak gets abruptly narrower with increasing system size. For ${L=256}$, the width of this peak may lie within the unsampled interval $0.099390< J< 0.099395$ (we do not have data for $J=0.0993915$ for this size).

\subsection{Evidence for $O(4)$--ordered regime}
\label{subsec:semicircle}

We now give evidence that at the longest scales we can access the system is in a regime which is described approximately by the $O(4)$ sigma model in its \textit{ordered} or Goldstone phase, as discussed in Sec.~\ref{sec:rgpicture}.

A first indication of this is that for $L\sim 128$ the order parameter values are fairly close to their extrapolated $L=\infty$ values (Sec.~\ref{subsec:firstorder}), while at the same time the moment ratios are close to the $O(4)$ values. 
This is consistent with being in the $O(4)$-ordered regime. 
It is not consistent with a standard critical regime (where the order parameters would be scaling to zero with $L$) \
and it is not consistent with a standard first order regime (where there would be no $O(4)$ symmetry).

\begin{figure}[t]
 \begin{center}
 \includegraphics[width=0.95\linewidth]{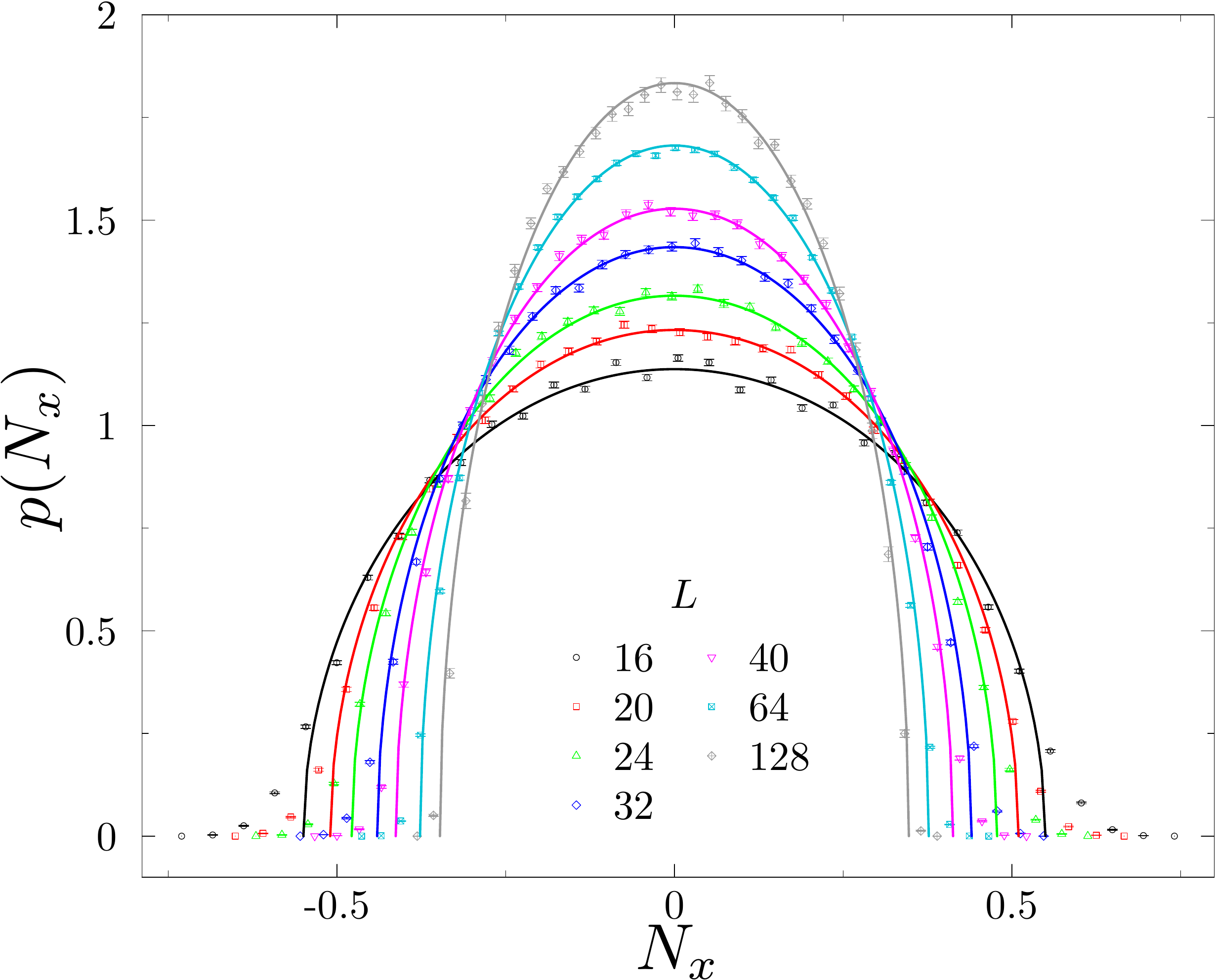}
 \end{center}
\caption{ Probability distribution of $N_{x}$ for several system sizes at $J=0.0993915$. Continuous lines are fits to Eq.~\ref{eq:semicircle} with $R$ as the only parameter (the fitted region is with $p(N_{x})>0.15$).
The extrapolated value of the order parameter in Eq.~\ref{eq:extrapolatedorderparams} corresponds to a semicircle with peak height of $2.0$.
}
\label{fig:probdistN}
\end{figure}

For a concrete check, we compare the probability distribution of a component of the order parameter with that expected for the sigma model in the ordered phase.
Deep in the ordered phase, the 4-component order parameter can be treated as spatially constant, and has a uniform probability distribution on the sphere
\be
|{\bf n}|^2 = R^2.
\ee
Integrating out 3 components gives the probability distribution for a single component, which is a semicircle:
\be\label{eq:semicircle}
P(n_1) = \f{2}{\pi R^2} \sqrt{ R^2 - n_1^2}.
\ee
This semicircle distribution is a hallmark of $O(4)$ order.

First, the probability distribution for $N_{x}$ is shown in Fig.~\ref{fig:probdistN} for several system sizes. We fit each curve to Eq.~\ref{eq:semicircle}, with $R$ as the only parameter (continuous lines). 
The extrapolated $L=\infty$ value of the order parameter  in Eq.~\ref{eq:extrapolatedorderparams} would correspond to the 
the height of the peak of the curve in Fig.~\ref{fig:probdistN} being $\sim 2.0$.

\begin{figure}[t]
 \begin{center}
 \includegraphics[width=0.95\linewidth]{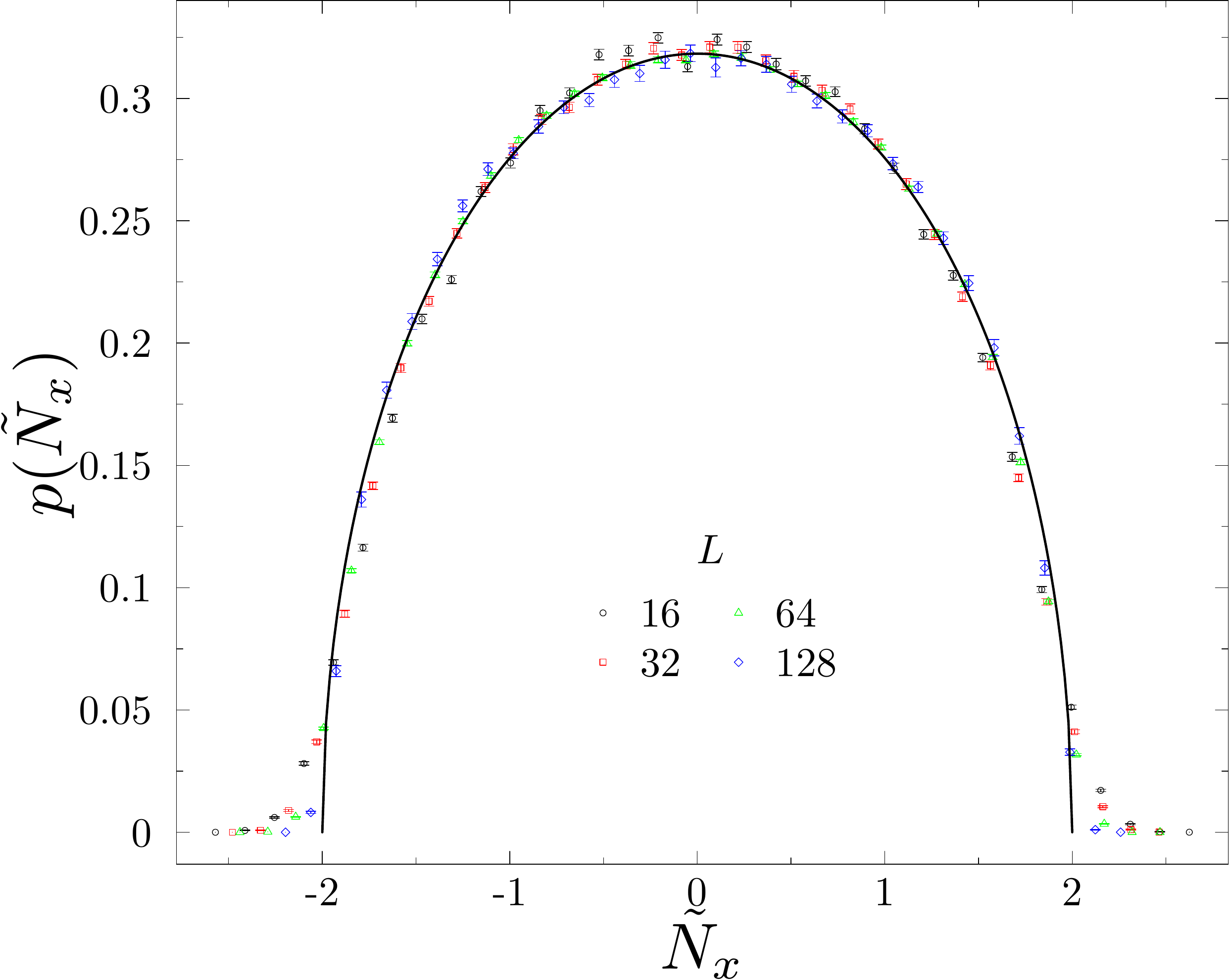}
 \includegraphics[width=0.95\linewidth]{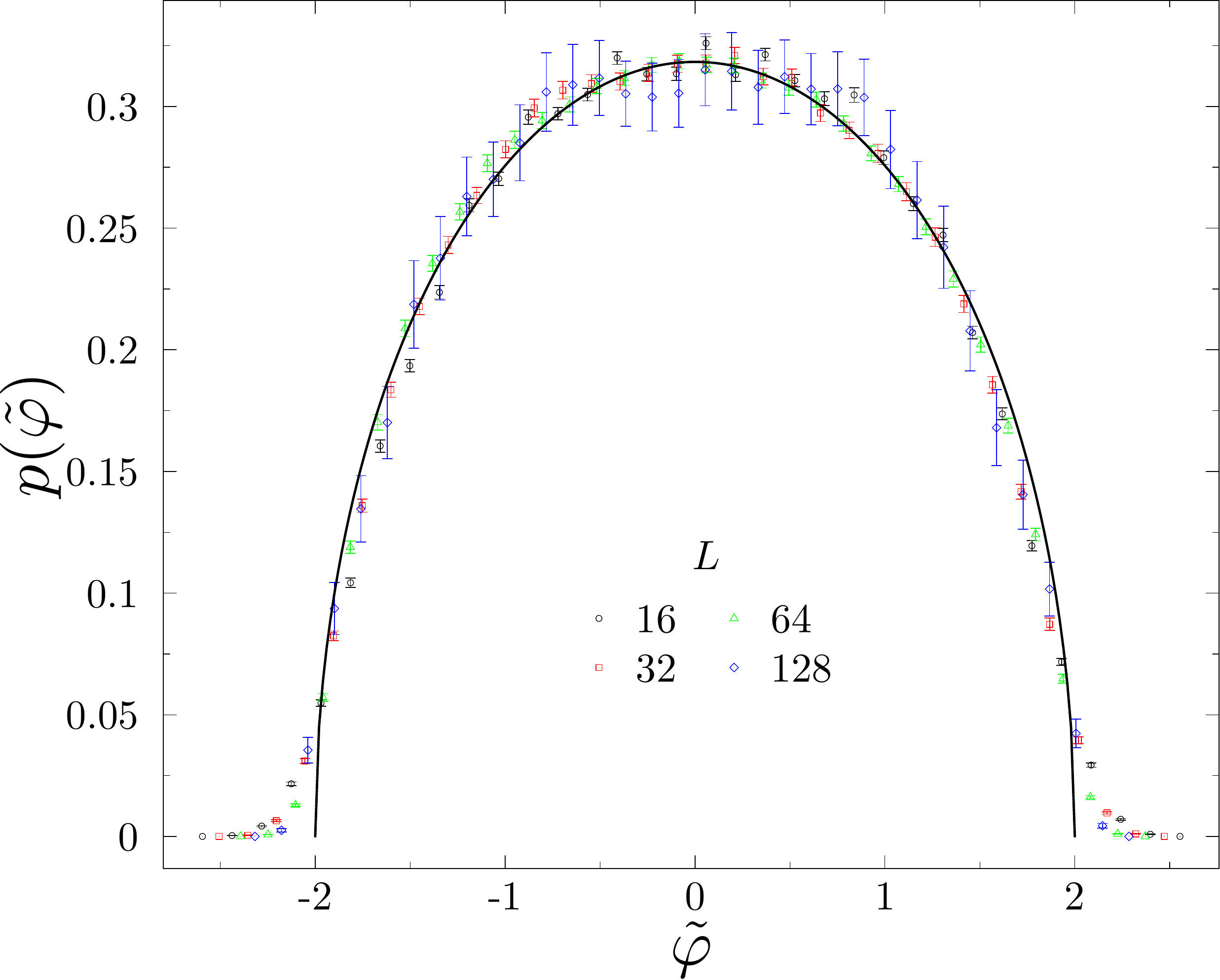}
 \end{center}
\caption{Probability distribution of the normalized $\tilde N = N_{x}/\sqrt{\<N_{x}^{2}\>}$ and $\tilde\varphi = \varphi/\sqrt{\<\varphi^{2}\>}$ for several system sizes. Curves show semicircle distribution with radius $R=2$.} 

\label{fig:probdistnorm}
\end{figure}

In Fig.~\ref{fig:probdistnorm} we show the probability distributions for the {normalized} order parameter components, $\tilde N_x$ and $\tilde \varphi$ with variance 1. 
We see that the shape of the probability distribution changes more weakly than the overall width. 
The expected semicircle form is shown as a black line (without any fitting parameter, since the variance is now fixed).
Error bars are smaller for $N_x$ than for $\varphi$.

While deviations from the semicircle are apparent for the smaller sizes, 
the distributions 
 for $L=64$ and $L=128$ match extremely  well to the semicircle.
The amount of weight in the tails $|\tilde N_x|>2$ is clearly decreasing as $L$ is increased, as expected from the semicircle form which has a bounded support.

Finally, for a more quantitative analysis, we compare cumulants of the order parameters with the values expected from the symmetric $O(4)$ sigma model in its ordered phase. 
Consider the cumulants (recall that $\tilde N_x$ is normalized to have unit variance)
\ba
C_4 & = \<\tilde N_x^4\> - 3 \<\tilde N_x^2\>^2, \\
C_6 & = \<\tilde N_x^6\> - 15 \<\tilde N_x^4\> \<\tilde N_x^2\> + 30 \<\tilde  N_x^2\>^3,
\end{align}
and similarly for $\tilde \varphi$.
In the $O(4)$ ordered regime, we expect from Eq.~\ref{eq:semicircle}:
\ba
C_4 & = -1, & C_6 & = 5,
\end{align}
for both $N_x$ and $\varphi$.
These values are very different from those that would be obtained for a Gaussian distribution (zero) or at a conventional first order transition.\footnote{At a conventional first order transition between N\'eel and VBS there is a probability $p(J,L)$ for the system to be in the VBS phase and $1-p(J,L)$ to be in the N\'eel phase,
with $p$ determined by the free energy difference between the two phases in the finite system (here $L$ is large).
Using the values in the N\'eel phase (from averaging over the sphere) and the VBS phase ($\varphi^2\sim \text{const}.$) gives, for a component of  $\vec N$, ${C_4^{(N)}=(3/5)(5p-2)/(1-p)}$ and ${C_6^{(N)} = (3/7) (16-77p+70p^2)/(1-p)^2}$, and for $\varphi$, ${C_4^{(\varphi)}=(1-3p)/p}$ and ${C_6^{(\varphi)} = (30p^2-15p+1)/p^2}$.  Strictly at $J_c$, the free energy densities of the two phases are equal, but in a finite system with periodic boundary conditions the N\'eel phase has an $O(\ln L)$ entropy associated with continuous symmetry breaking. As a result $p\rightarrow 0$ as a power law at large $L$. In this limit ${C_4^{(N)}=-6/5}$,  ${C_6^{(N)}=48/7}$,  and ${C_4^{(\varphi)}}$ and ${C_6^{(\varphi)}}$ both diverge.} 

{In Fig.~\ref{fig:Cs} we show these cumulants at $J=J_{c}$
as a function of the inverse of the system size, for both order parameters. 
Both $C_{4}$ and $C_{6}$ are relatively close to the values expected from the ordered sigma model even for the smaller sizes, and for the larger sizes they are very close.
This suggests  that the  system at $J_c$ is well described by the ordered phase of the symmetric sigma model for $L\sim 128$ and even $L\sim 64$, despite the fact that the ordered moment at $L=64$ is still quite far from its extrapolated value (Fig.~\ref{fig:Oatjc}).

\begin{figure}[t]
 \begin{center}
 \includegraphics[width=0.95\linewidth]{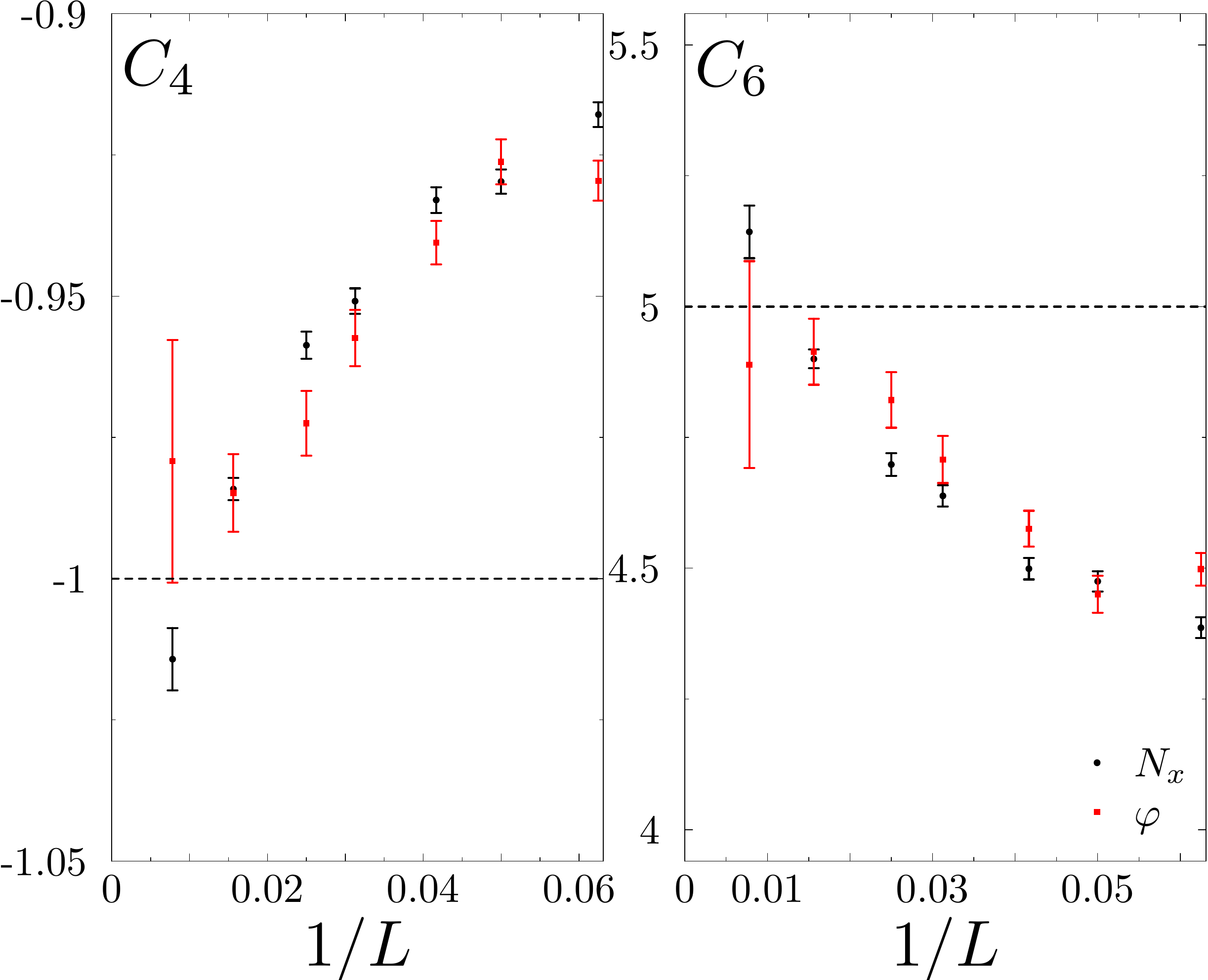}
 \end{center}
\caption{ Cumulants $C_4$ and $C_{6}$ at $J=0.0993915$ as function of the inverse of the system size, $1/L$. Dashed lines are the expected values for the $O(4)$-symmetric ordered sigma model, $C_{4}=-1$ and $C_{6}=5$.
}
\label{fig:Cs}
\end{figure}

\pagebreak

\section{Outlook}
\label{sec:conclusion}

We have exhibited a phase transition with emergent $O(4)$ symmetry 
and a regime where the $O(4)$ vector is effectively ordered.
This symmetry is only expected to be approximate,
and therefore the description using the long-range ordered $O(4)$ sigma model 
 is also only approximate even in the relevant range of length scales.
However, the usefulness of this concept in organizing our understanding of this phase transition is clear. 
We have shown that it allows us to predict various `quasiuniversal' features with good accuracy. 
As a zeroth order description for length scales of order, say, $100$, the `$O(4)$ spin flop' description is clearly far more useful than standard expectations for first order transitions, which are strongly violated because of the effective   continuous symmetry relating the two  phases at coexistence. 

The description of the data using the sigma model description could be further improved by considering Goldstone fluctuation corrections, as well as the leading weak anisotropies in the Lagrangian. Many other quantities, such as two-point functions or off-critical scaling forms, could also be understood with this approach.

The phenomenon we have argued for in the present model is likely to be more general  \cite{wang2017deconfined}. Examples may exist with a range of emergent symmetry groups $G$, and for a given $G$, there will be many different possibilities for the (smaller) microscopic symmetry group.

The present $O(4)$ regime may also appear in several other contexts.
From the renormalization group picture one would naively expect that at large scales the `easy-plane' N\'eel-VBS transition  \cite{qin2017duality,zhang2018continuous,kukloveasyplane, kragseteasyplane, kauleasyplane1, kauleasyplane2}  behaves similarly to the model discussed here \cite{tsmpaf06, metlitski2017intrinsic, wang2017deconfined}. 
(The easy plane transition involves competition between a \textit{two}-component N\'eel order parameter and a two-component VBS order parameter.)
However, recent work has argued that this transition can be continuous rather than first order \cite{qin2017duality,zhang2018continuous}.
This should be understood better.
 The classical 3D dimer model, with appropriate anisotropy, is another platform that could be used to study the physics of the easy-plane $\nccp^1$ model  \cite{Chenetal, sreejith2018emergent}.

The regime demonstrated here should also be looked for in two-flavour QED$_3$, and is an alternative possibility to flow to a scale-invariant fixed point argued for previously on numerical grounds  \cite{qedcft}. (The effect of $O(4)$-breaking perturbations within the $O(4)$--ordered phase is expected to be much weaker in that  context \cite{wang2017deconfined}.)
  
 A long-range-ordered $SO(5)$ regime may occur for deconfined criticality with 5 components, for example $SU(2)$-symmetric magnets at the N\'eel-VBS transition on the square lattice  \cite{wang2017deconfined}. 
Existing data shows features that seem  compatible with a pseudocritical regime, for example drifting exponents that at large size tend to values that are incompatible with the exponents of a conformal fixed point \cite{DCPscalingviolations,SimmonsDuffinSO(5),Nakayama,PolandReview}, at least if conventional finite-size scaling is assumed.\footnote{In the presence of an appropriate dangerously irrelevant variable, the assumption of finite size scaling can be invalid  \cite{DCPscalingviolations, sandvik2parameter}.}
However if the flow is eventually to the long-range-ordered state,  the lengthscale needed to see this regime is very large and appears to be beyond the reach of current simulations \cite{DCPscalingviolations}. Indeed on the presently accessible lengthscales the $SO(5)$ symmetry in various models resembles an exact symmetry of the infrared theory 
\cite{emergentso5,sreejith2018emergent}. 

In the future it will be useful to quantify more precisely the the accuracy of the emergent approximate $O(4)$ symmetry  in the model studied here as a function of $L$.
A more careful comparison with expectations from the ordered sigma model, taking finite size effects into account, as well as the effects of finite ${J-J_c}$, 
would also be enlightening. Simulations at larger $L$ might be able to probe the eventual flow away from the symmetric sigma model (Fig.~\ref{fig:deltajflows}).

Finally, here we have focussed only on the critical point at $J_A=J_c$, $J_B=0$. 
It will also be important to investigate the transition both in the presumably more strongly first-order regime at negative $J_B$ and in the presumably more weakly first-order regime at larger $J_B$.
In Sec.~\ref{sec:rgpicture} we noted an alternative scenario in which  approximate $O(4)$ arises because the model is by accident fine-tuned close to an unstable $O(4)$--invariant \textit{tricritical} point. 
We have argued this is unlikely as it would require fine-tuning 
(and an appropriate such fixed point may not exist).
This could be checked directly by checking that the $O(4)$ ordered regime is robust against small changes of the microscopic parameters.

\vspace{2pt}
\emph{Related Work:} After completion of this work we became aware of a numerical study of a deformed $JQ$ model by Zhao et al. \cite{zhao2018symmetry}. That model also has a transition with 4 order parameter components, and the authors also find a first order transition with emergent $O(4)$ symmetry, as here.

\acknowledgements
The deformation of the loop model studied here was  suggested to one of us (PS) by Anders Sandvik and Ribhu Kaul, before the advent of emergent symmetries, as a way to interpolate between the DCP in the symmetric model and a strongly first order transition.
We thank them for discussions.
We thank G. J. Sreejith and S. Powell for discussions and collaboration on  recent related work, and J. Chalker for discussions. 
PS acknowledges Fundaci\'on S\'eneca grant 19907/GERM/15.
AN acknowledges EPSRC Grant No.~EP/N028678/1.

\appendix
\section{Comparison of two VBS components}
\label{sec:twovbs}

\begin{figure}[b]
 \begin{center}
 \includegraphics[width=0.95\linewidth]{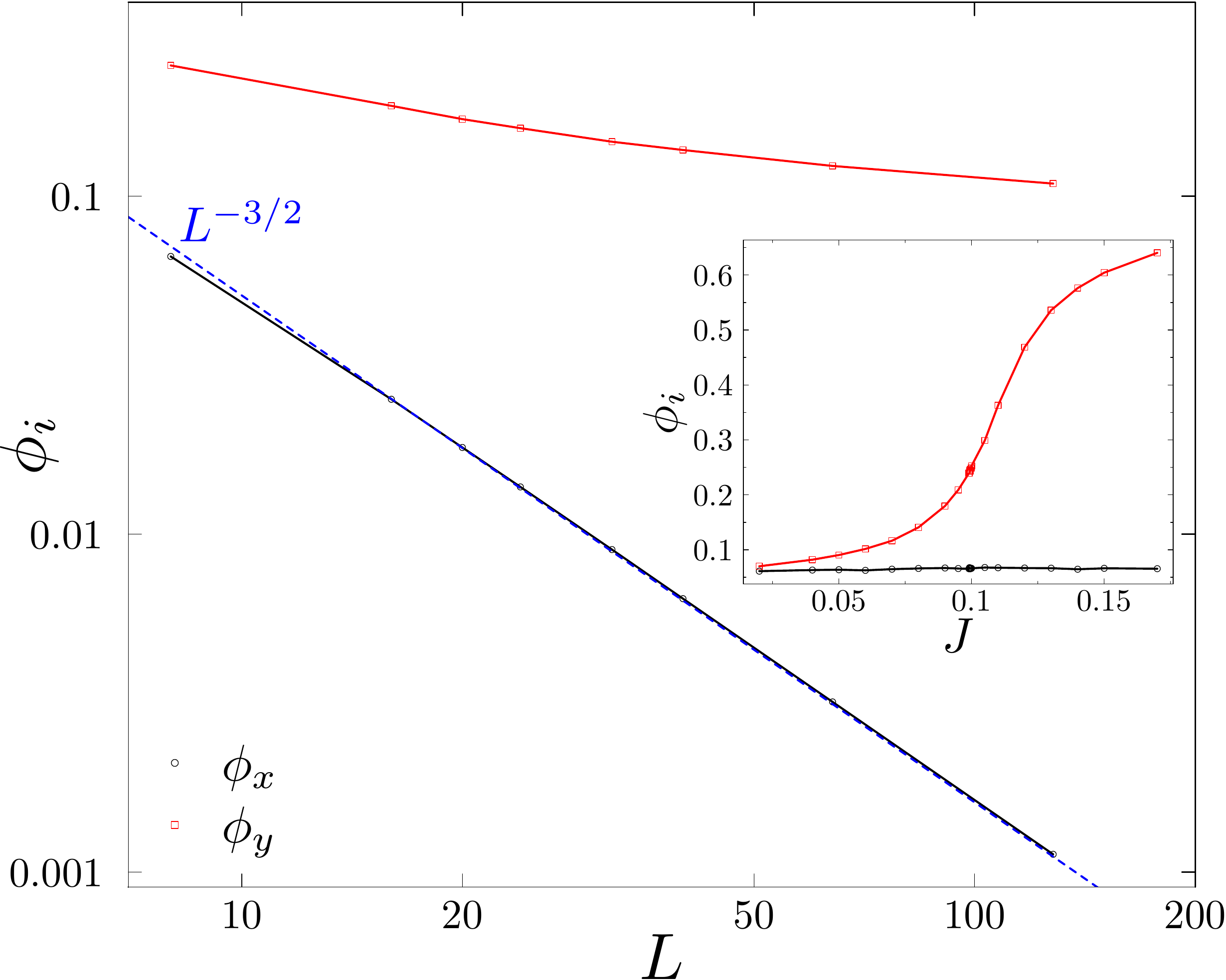}
 \end{center}
\caption{ 
Main panel: we show the two components of $\phi$ (see Sec.~\ref{sec:model}) as a function of system size at $J=0.0993915$ (close to $J_c)$. Note the log-log scale. The definition is ${\phi_i = \sqrt{\< \phi_i^2\>}}$, where the operator inside the expectation value is averaged over the full system volume.
First, observe that for our choice of couplings (Eq.~\ref{eq:jajb})
there is a very strong asymmetry between $\varphi=\phi_x$ and $\phi_y$ even at small scales. 
The magnitude of $\phi_y$ decreases as $L^{-3/2}$ (dashed line), as expected for a \textit{massive} field.
This shows that the $O(4)$ symmetry we find is not just a consequence of proximity to the $SO(5)$ invariant critical (or pseudocritical) point in the symmetric model with ${J_A=J_B}$.
Second, note that the log-log plot for $\varphi=\phi_x$ (red line) is not straight, showing that the power law scaling that would be expected at a critical point does not apply here. Instead, $\varphi$ extrapolates to a finite value as $L\rightarrow \infty$ (Sec.~\ref{subsec:firstorder}).
Inset: we show the two components of $\phi$  as a function of $J$ in a small system ($L=8$). Again the strong asymmetry between $\varphi=\phi_x$, which orders at the transition, and $\phi_y$, which remains massive, is apparent.
}
\label{fig:twophis}
\end{figure}

In Fig.~\ref{fig:twophis} we show the two components of the VBS order parameter in Eq.~\ref{eq:phiuniform}, $\phi_x$ and $\phi_y$, both as a function of $L$ at $J_c$, and as a function of $J$ for $L=8$. 
The figure shows that at the transition $\phi_y$ is a massive field with a short correlation length, and that there is a strong asymmetry between $\phi_x$ and $\phi_y$ even at short scales. 
This confirms that the coupling values we have chosen (i.e. $J_B=0$) are sufficiently far from the symmetric critical point with $J_A=J_B$.
The symmetric critical point has very accurate $SO(5)$ symmetry. 
If we studied a weakly asymmetric model (i.e. $J_A-J_B$ too small) then there would be an intermediate range of scales with approximate $SO(5)$, and  we would see $O(4)$ as a trivial consequence of this. It is clear from Fig.~\ref{fig:twophis} that this is not the explanation for the $O(4)$ that we see.

\bibliographystyle{unsrt}
\bibliography{spinfloprefs}

\end{document}